\documentclass[superscriptaddress,showpacs]{jfm}
\usepackage{natbib}
\usepackage{epsfig}
\usepackage{graphicx}
\usepackage{amssymb}
\usepackage{amsmath}
\usepackage{citesort}
\def\be{\begin{eqnarray}}
\def\ee{\end{eqnarray}}
\def\beq{\begin{eqnarray}}
\def\eeq{\end{eqnarray}}
\begin{document}

\title{Controlled impact of a disk on a water surface: Cavity dynamics.}
\author[R. Bergmann {\it et al.}]{Raymond Bergmann$^{1,2}$, Devaraj van der Meer$^1$, Stephan Gekle$^1$, Arjan van der Bos$^1$, and Detlef Lohse$^1$}
\affiliation{$^1$ Department of Applied Physics and J.M. Burgers
Centre for Fluid Dynamics, University of Twente, P.O. Box 217,
7500 AE Enschede, The Netherlands\\
$^2$ Department of Physics and Center for Fluid Dynamics, The
Technical University of Denmark, DK-2800 Kgs. Lyngby, Denmark}
\date{\today}
\setcounter{page}{1}

\maketitle

\begin{abstract}
In this paper we study the transient surface cavity which is created
by the controlled impact of a disk of radius $h_0$ on a water
surface at Froude numbers below 200. The dynamics of the transient
free surface is recorded by high speed imaging and compared to
boundary integral simulations. An excellent agreement is found
between both. The flow surrounding the cavity is measured with high
speed particle image velocimetry and is found to also agree
perfectly with the flow field obtained
from the simulations.\\
We present a simple model for the radial dynamics of the cavity
based on the collapse of an infinite cylinder. This model accounts
for the observed asymmetry of the radial dynamics between the
expansion and contraction phase of the cavity. It reproduces the
scaling of the closure depth and total depth of the cavity which are
both found to scale roughly as $\propto \text{Fr}^{1/2}$ with a
weakly Froude number dependent prefactor. In addition, the model
accurately captures the dynamics of the minimal radius of the
cavity, the scaling of the volume $V_{bubble}$ of air entrained by
the process, namely $V_{bubble}/h_0^3 \propto (1 +
0.26\text{Fr}^{1/2})\text{Fr}^{1/2}$, and gives insight into the
axial asymmetry of the pinch-off process.
\end{abstract}

\section{Introduction}
A spectacular example of free surface flow is the impact of an
object on a liquid: The impact creates a splash and a transient
cavity. This surface cavity then violently collapses under the
influence of the hydrostatic pressure. At the singularity where the
walls of the cavity collide, two powerful jets develop, one
downwards and the other one upwards up to several meters high,
making this fast event an impressive scene. Research into the
physics of these transient surface cavities started at the beginning
of the twentieth century when A.M. Worthington published his famous
work "A study of splashes" (\cite{wor08}). His photographs revealed
a wealth of phenomena of unanticipated complexity (\cite{wor1897}).
Although much has been contributed to the understanding of these
phenomena, many of the intriguing questions posed by Worthington's
photographs resonate still today (\cite{rei93,fed04}).\\

All investigations since Worthington's studies entailed experiments
with a freely falling object impacting on the free surface. To gain
further insight into such impact events, we built a setup in which
we attach the impacting object to a linear motor. In this way we
gain full control over the impact velocity, which now turns from a
response observable into the key control parameter of the
system.\\

The dynamics of a surface cavity are of enormous practical
importance in many natural and industrial processes: Raindrops
falling onto the ocean entrain air (\cite{ogu90,ogu95,pro97}) and it
is this mechanism which is one of the major sinks of carbon dioxide
from the atmosphere. Droplet impact and the subsequent void collapse
are also a significant source of underwater sound (\cite{pro89}) and
a thorough understanding is therefore crucial in sonar research.
High speed water impacts and underwater cavity formation are
moreover of relevance to military operations
(\cite{Gil48,lee97,Duc06}). In the context of industrial
applications, drop impact and the subsequent void formation are
crucial in pyrometallurgy (\cite{liow96,mor00}), in the food
industry, and in the context of ink-jet printing
(\cite{Le98,che02,jon06a,jon06b}). A similar series of events as in
water can even be observed when a
steel ball impacts on very fine and soft sand (\cite{Tho01,loh04a,Roy05,cab07}).\\

Although in some of the literature the deceleration of the
impacting body was minimized by choosing the properties of the
body such that the velocity of the impactor remained roughly
constant during the time the cavity dynamics were observed
(\cite{gla96,gau98}), the velocity of the body nevertheless
remained a response parameter set by the system. Our use of a linear
motor to accurately control the position, velocity, and
acceleration of the impacting object constitutes the key
difference between our
work (see also \cite{ber06,gek08}) and all previous literature.\\

In this article, we will use observations from experiments and
boundary integral simulations to construct a model which
accurately describes the radial dynamics of the cavity.\\
In Section \ref{sec:experiment} we present results from our
controlled experiment and compare them to the boundary integral
simulations. More specifically, in
subsection~\ref{sec:exp_interface} we discuss the dynamics of the
free surface and continue in subsection \ref{sec:exp_flow} with the
topology and magnitude of the flow surrounding the cavity obtained
by particle image velocimetry.\\
In Section~\ref{sec:model2} we will derive a model which captures
the radial dynamics of the cavity. We will use the model to
investigate the following key characteristics of the transient
surface cavity: First, the depth at which the pinch-off will occur
is discussed in subsection \ref{sec:closure depth}. Then, in
subsection \ref{sec:air entrainment} the amount of air entrained
by the cavity collapse is studied.  
The article is concluded in Section~\ref{sec:conclusion}. The
results from our earlier paper \cite{ber06} that are relevant to the
present study are reviewed in Appendix A, together with some
additional information on the time evolution of the neck radius and
the cavity. Finally, Appendix B discusses the dynamics of the
minimal radius within the context of the model.

\section{Experimental and numerical
results} \label{sec:experiment}

\begin{figure}
\begin{center}
\includegraphics[width=0.7\textwidth]{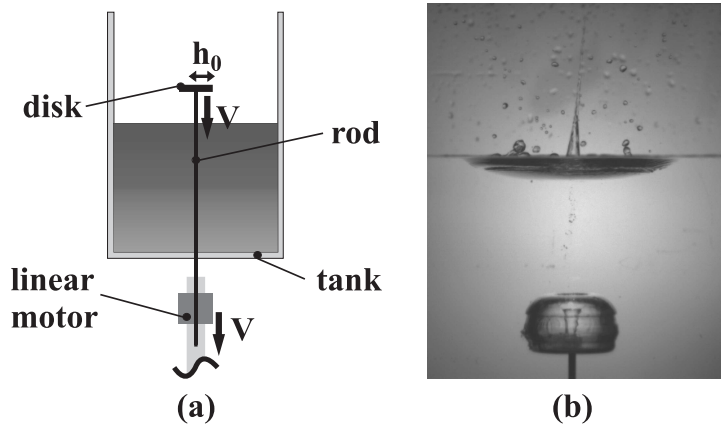}
\caption{(a) Schematic figure of the experimental setup with the
linear motor controlling the rod and disk. The vertical rod runs
through a seal in the bottom of the glass water tank and is pulled
down by the linear motor in order to impact the disk on the water
surface. Here, $h_0$ is the radius and $V$ the controlled and
constant velocity of the disk. (b) The formation of an upward and
downward jet after the cavity has closed. In the present study we
will focus on the cavity dynamics until pinch-off, just before the
jet formation.} \label{fig:experimental_setup}
\end{center}
\end{figure}

\subsection{Experimental setup and procedure}

A sketch of the setup is seen in
Fig.~\ref{fig:experimental_setup}a. A disk of radius $h_0$ is
mounted on top of a thin rod ($\varnothing$ 6 mm). This rod runs
through a seal in the bottom of a large tank
(500~mm$\times$500~mm$\times$1000~mm) and is connected at the
lower end to a Thrusttube linear motor which is used to determine
and control the velocity and acceleration of the disk. The
position of the motor (and thus of the disk) along the vertical
axis is measured with a spatial accuracy of 5 $\mu$m over a range
of 1 m, the large acceleration of the motor (up to 30 $g$, with
$g$ the gravitational acceleration) makes it possible to perform
impact experiments with
constant velocities up to 5 m/s.\\

The effect of the small diameter of the rod on the global flow and
dynamics of the cavity is assumed to be negligible. As the minimum
radius for the disk used in the experiments is 10 mm, the ratio of
the cross--sectional area of the rod and the surface of the disk
is always smaller than 9\%. Since the rod is mounted in the center
of the disk, where stagnation would normally occur, the influence
on the radially outward flow below the disk
is presumably small.\\

Using the flat plate approximation, we can also estimate the
direct contribution of the boundary layer of the rod to the axial
flow. The boundary layer thickness $\delta$ for a flat plate is
given by Blasius' solution $\delta \approx 5\sqrt{\nu \Delta t}$,
where $\nu$ is the kinematic viscosity of water and $\Delta t$ the
time the boundary layer has to develop. We will equate the time
$\Delta t$ to the duration of the experiment, namely to the time
interval starting from the impact of the disk until the collapse
of the void, which in our experiments is found to scale as $\Delta
t \approx 2.5 \sqrt{h_0/g}$, as will be discussed in detail in
subsection \ref{sec:closure depth}. For the largest disk size in
the experiment $h_0 = 40$ mm, this result predicts a maximum
boundary layer thickness of 1.8 mm. Under most experimental
conditions of this study it is considerably thinner.\\

In our experiments we pull the disk down with a {\it constant}
velocity $V$. Making this main control parameter dimensionless, we
obtain the Froude number $\text{Fr} =V^2/(gh_0)$. The liquid
properties are expressed in terms of the Reynolds number $\text{Re}
=Vh_0/\nu$ and the Weber number $\text{We} =\rho V^2h_0/\sigma$,
where  $\sigma$ denotes the surface tension and $\rho$ the fluid
density. Since the Reynolds number and the Weber number are
considerable on the large scales of
Fig.~\ref{fig:experimental_setup}, the viscosity and the surface
tension do not seem to play a role. To be more precise, in our
experiment the Reynolds number ranges between 500 and
$1.6\cdot10^{4}$ and the Weber number ranges between 34 and
$8.8\cdot10^3$. Note however that under only slightly different
conditions, namely replacing the disk by a cylinder submerged in
water to avoid the splash, capillary waves {\it do} play a role (see
\cite{gek08}). For the impact of a disk we find the only important
dimensionless parameter to be the Froude number, i.e., the ratio of
kinetic to gravitational energy, which ranges from $0.6$ to $200$ in
our experiments. It is convenient to use the amount of time $\tau$
remaining until cavity collapse which is given by $\tau=t-t_{coll}$
with $t_{coll}$ the collapse time.\\

\subsection{Numerical method}

The numerical calculations are performed using a boundary integral
method (\cite{pro02,pow95,ogu93}) based on potential flow. This
assumption excludes viscous effects and vorticity, which due to the
short duration of the phenomenon
and the high Reynolds number seems reasonable.\\
Our code uses an axisymmetric geometry thus reducing the surface
integrals to one-dimensional line integrals. For the time-stepping
an iterative Crank-Nicholson scheme is employed. The size of each
time step is calculated as $t_{step}=f\cdot\min(t_{node})$ with
$t_{node}=\Delta s/v_{node}$, where $\Delta s$ is the distance to
the neighboring node and $v_{node}$ the local velocity. With the
safety factor $f$ chosen to be 5\% this procedure reliably prevents
collisions of two nodes
which would lead to serious disturbances in the numerical scheme.\\
The number of nodes is variable in time, with the node density at
any particular point on the surface being a function of the local
curvature. This procedure guarantees that in regions with large
curvatures, especially around the pinch-off point, the node density
is always high enough to resolve the local details of the surface
shape. At the same time, no computation power is wasted on an
exceedingly high node density in flat regions towards infinity
(which in our simulations is chosen to be 100 disk radii away
from the central axis). To avoid numerical disturbances, we
employ a regridding scheme in which at every second time-step
the surface nodes are completely redistributed placing the new
nodes exactly half-way between the old nodes.\\

A particularly sensitive issue is the modeling of the crown splash
created when the disk impacts the water surface. After first
shooting upwards in a ring shape, the splash quickly breaks up
into a large number of drops (which are ring-shaped due to the imposed
axial symmetry). These drops do not further influence the cavity
behavior and therefore need not be accounted for in our numerical code.
In most simulations presented in this work, the crown-splash evolves
normally until drop pinch-off. As this happens, the surface is
reconnected at the pinch-off location and the drop is discarded.\\

\subsection{Interface} \label{sec:exp_interface}

\begin{figure}
\begin{center}
\includegraphics[width=0.75\textwidth]{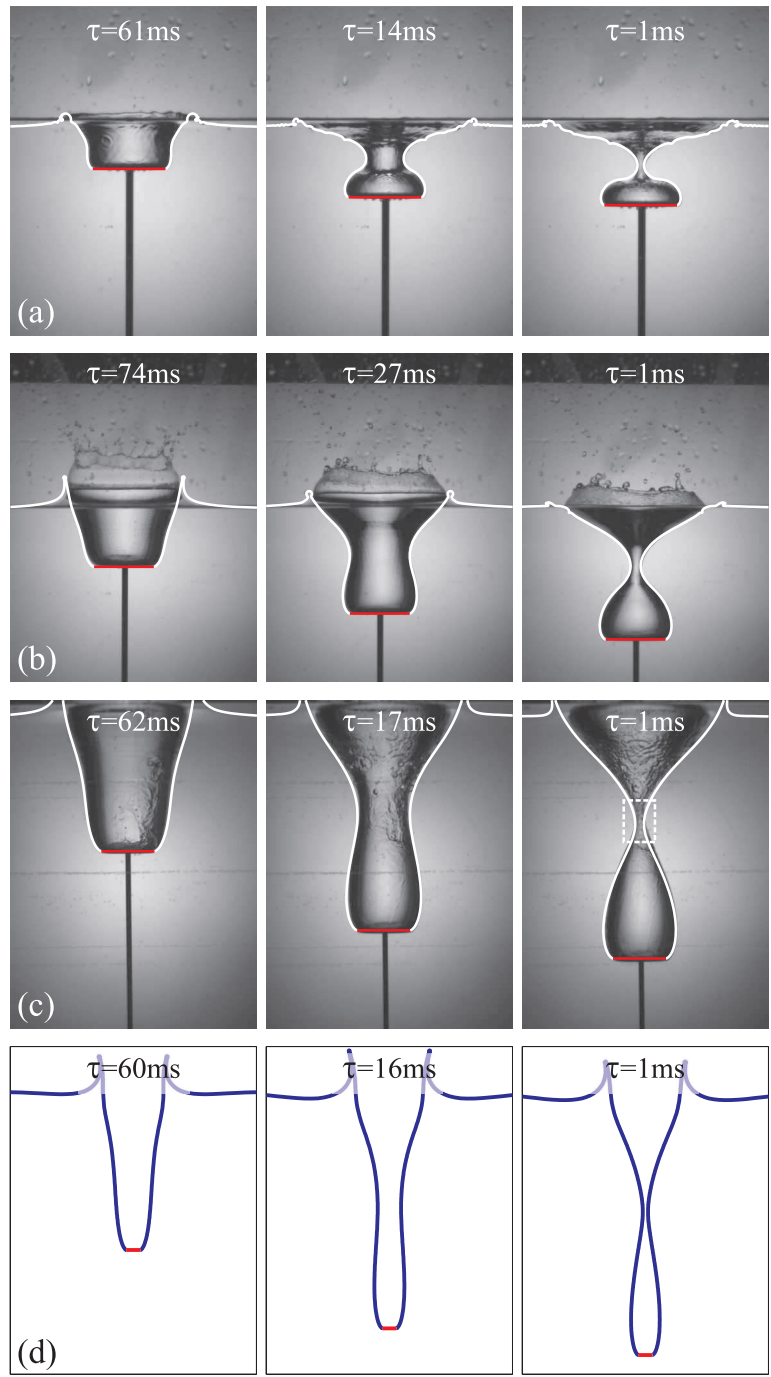}
\caption{Snapshots of the formation and collapse of a surface void
in the plunger experiment: A linear motor pulls down a disk of
radius $h_0$ = 30 mm through the water surface at a constant
velocity (a) $V = 0.5$ m/s (Fr = 0.85); (b) $V = 1.0$ m/s (Fr =
3.4); (c) $V = 2.0$ m/s (Fr = 13.6); and (d) Fr = 200. The collapse
of the void is imaged at 1000 frames per second. The lines (overlay)
are the void profiles obtained from boundary integral simulations.
Without the use of any free parameter, neither in time nor in space,
an excellent agreement between the simulation and experiment is
found in (a) and (b). Due to a (mild) {\it surface seal} there is a
discrepancy between the simulations and the experiment in sequence
(c), both in the top region near the splash and in the pinch-off
region. The region of the dashed box is shown enlarged in
Fig.~\ref{fig:comparison_zoom}. The experimental data for (d) (not
shown) is severely dominated by a strong surface seal, which is an
air effect and not the focus of the present study.} 
\label{fig:comparison}
\end{center}
\end{figure}

\begin{figure}
\begin{center}
\includegraphics[width=0.23\textwidth]{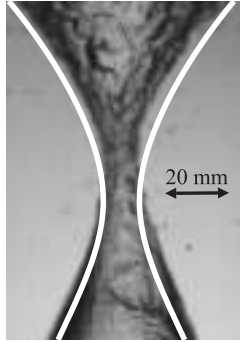}
\caption{Enlargement of the region around the pinch off point at
$\tau =$ 1 ms from the sequence Fig.~\ref{fig:comparison}c (Fr =
13.6). A significant discrepancy can be seen for the depth of the
pinch--off between the boundary integral simulation (white line) and
the experimental recording. The origin of this discrepancy is the
airflow in the cavity as will be elaborated elsewhere.}
\label{fig:comparison_zoom}
\end{center}
\end{figure}

\begin{figure}
\includegraphics[width=1\textwidth]{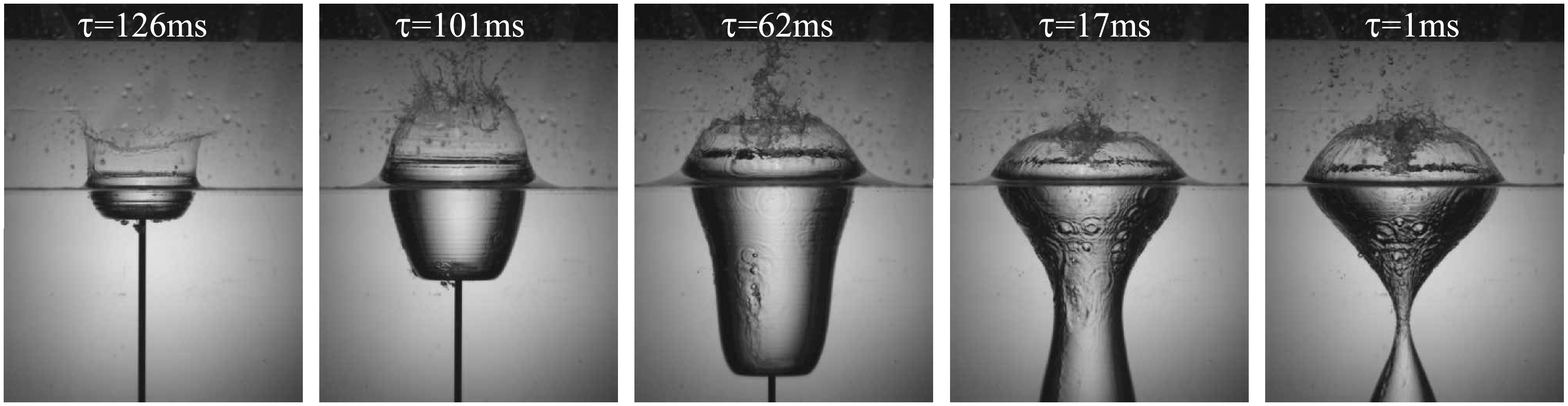}
\caption{Snapshots of the surface seal which occurs for a disk of
radius $h_0$ = 30 mm impacting the water surface at a constant
velocity of $V = 2.0$ m/s (Fr = 13.6), i.e., under the same
conditions as Fig.~\ref{fig:comparison}c.} \label{fig:surface_seal}
\end{figure}

The series of events typical for the experimental range of
$1<\text{Fr}<100$ is seen in the snapshots of
Figs.~\ref{fig:comparison}a, b, and c. Upon impact a splash, an
outward moving crown of water, is formed. A void is created which
collapses due to the hydrostatic pressure and a singularity arises
when the collapsing walls of the void collide with each other. Two
jets emerge in this experiment: One upwards straight into the air,
and one downwards into the
entrained air bubble (see Fig.~\ref{fig:experimental_setup}b).\\

In each of Figs.~\ref{fig:comparison}a, b, and c the experimental
sequence is overlaid with the results of our boundary integral
simulation. For Fr=0.85 and Fr=3.4 (Figs.~\ref{fig:comparison}a and
b), the cavity dynamics is found to be captured extremely well by
the numerical result, represented by the overlaid lines. Note that
this is a one-to-one comparison between simulation and experiment,
without any rescaling in time or space. Due to the axisymmetry of
the code it is not possible to capture the full details of the
splash and since our focus is on the cavity dynamics we chose to
simply take out any droplets which are released from the splash.
Surface tension however still expresses itself in small capillary
waves in the region of the splash. These waves are most notable in
Fig.~\ref{fig:comparison}a. As was mentioned before, similar
capillary waves (but from a different origin) are found to have a
significant influence on the closure of the cavity for a submerging
cylinder (\cite{gek08}). For the impacting disk discussed in this
paper however they do not affect the closure.\\

The results for Fr=0.85 in Fig.~\ref{fig:comparison}a illustrate the
effect of the relative importance of gravity. In the last frame of
Fig.~\ref{fig:comparison}a it can be seen that the cavity is less
symmetric in axial direction around the closure point compared to
the experiment performed at Fr=3.4 shown in
Fig.~\ref{fig:comparison}b. In the third sequence at Fr=13.6
(Fig.~\ref{fig:comparison}c), which goes beyond the experimental
Froude number range described in \cite{ber06}, significant
deviations between the experiment and the numerical cavity shape are
found, most notably in the enlargement of
Fig.~\ref{fig:comparison_zoom} at the depth of the cavity closure.
The closure of the cavity is found to be somewhat deeper in the
numerics as compared to the experiments. This deviation can be
attributed to a secondary effect due to the surrounding air, called
the {\it surface seal} (see Fig.~\ref{fig:surface_seal}). This
phenomenon was first reported by \cite{wor08} and later investigated
in more detail by \cite{Gil48}. Note that the impact experiment of
Fig.~\ref{fig:surface_seal} is performed under the same conditions
as Fig.~\ref{fig:comparison}c. The surface seal is the entrainment
of the initially outward moving splash by the air rushing into the
expanding cavity. If the airflow is strong enough, the splash will
close on the axis of symmetry and completely seal off the top of the
cavity above the height of the
undisturbed water surface.\\

The surface seal is found to become more pronounced at higher impact
velocity, where the surface seal occurs earlier and more liquid is
involved in this closure. Accordingly, there is also a larger
influence on the shape of the cavity at higher impact velocity.
Since this article aims to treat the purely pressure driven collapse
of the cavity, without the contributions of the surrounding air, our
experimental range is limited by the occurrence of the surface seal.
In the simulations we therefore intentionally do not include the
air. This explains the discrepancy of Fig.~\ref{fig:comparison}c
(enlarged in Fig.~\ref{fig:comparison_zoom}), since contrary to the
experiments, no surface seal occurs in the numerics due to the
absence of air. In Fig.~\ref{fig:comparison}d we go far beyond the
experimentally available range by performing simulations at a
Froude number of 200.\\

It is instructive to compare the present boundary integral
simulation results with those reported by \cite{gau98}, who reported
a bulging contraction of the cavity at the surface level. He found
this contraction to close for $\text{Fr}\geq200$ and interpreted it
as a surface seal in the absence of air. We found no evidence for
such a surface seal in our simulations, even for considerably larger
Froude numbers, and surmise that the effect reported by \cite{gau98}
may be connected to using an insufficient number of nodes in the
splash region caused by the
limited amount of computational power available at that time.\\

\subsection{Flow field} \label{sec:exp_flow}

\begin{figure*}
\includegraphics[width=1\textwidth]{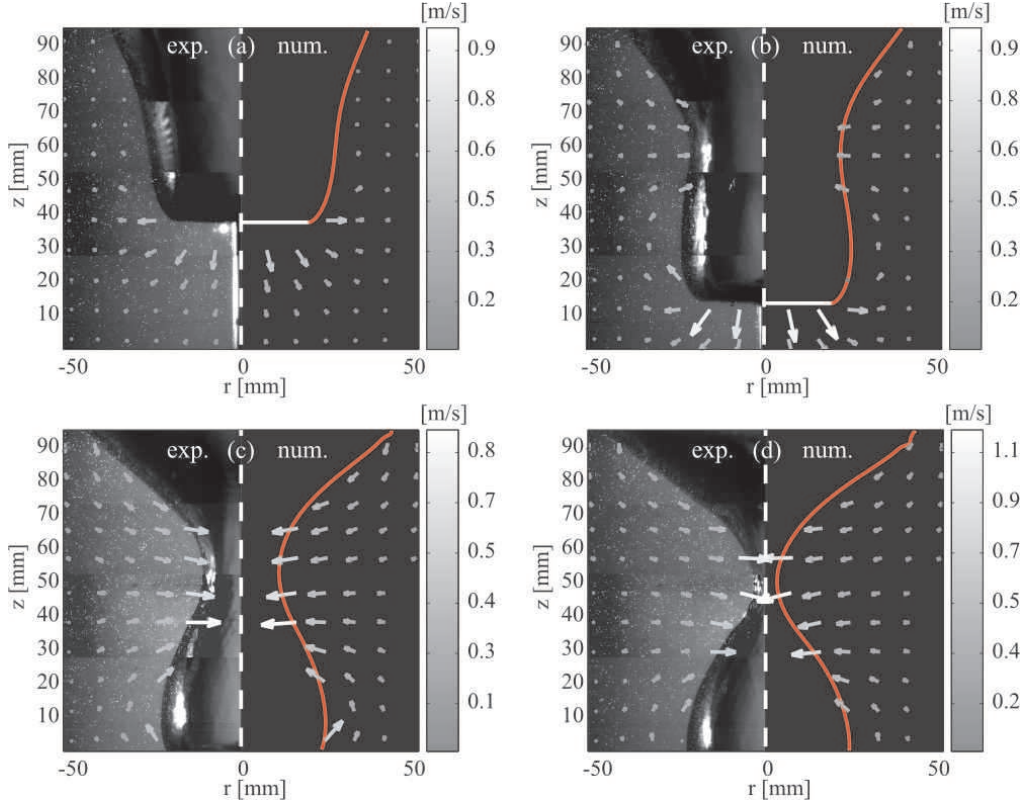}
\caption{Comparison of the flow field obtained from experiments and
boundary integral simulations for a disk of radius $h_{disk}$ = 20
mm which impacts with a velocity of $V = 1.0$ m/s (Fr = 5.1). The
figures show the flow field at $\tau=$ (a) 49 ms; (b) 25 ms; (c) 7
ms; and (d) 1 ms. The left side of each image shows the flow field
(overlaid vectors) obtained from the experiment by four high-speed
PIV recordings at 6000 frames per second. The four separate
recordings were taken at different depths and combined to give the
flow field at high resolution. The recordings on the left side of
each image also illustrate the degree of reproducibility of the
experiment, as the match between the four PIV recordings at
different depths obtained from four repetitions of the same
experiment (see main text) is perfect. For clarity only 0.7\% of the
measured vectors is shown. The right side of each of the images
shows the void profile and the corresponding flow field (overlaid
vectors) obtained from the boundary integral calculations.
} \label{fig:PIV global}
\end{figure*}

\begin{figure}
\centering
\includegraphics[width=0.9\textwidth]{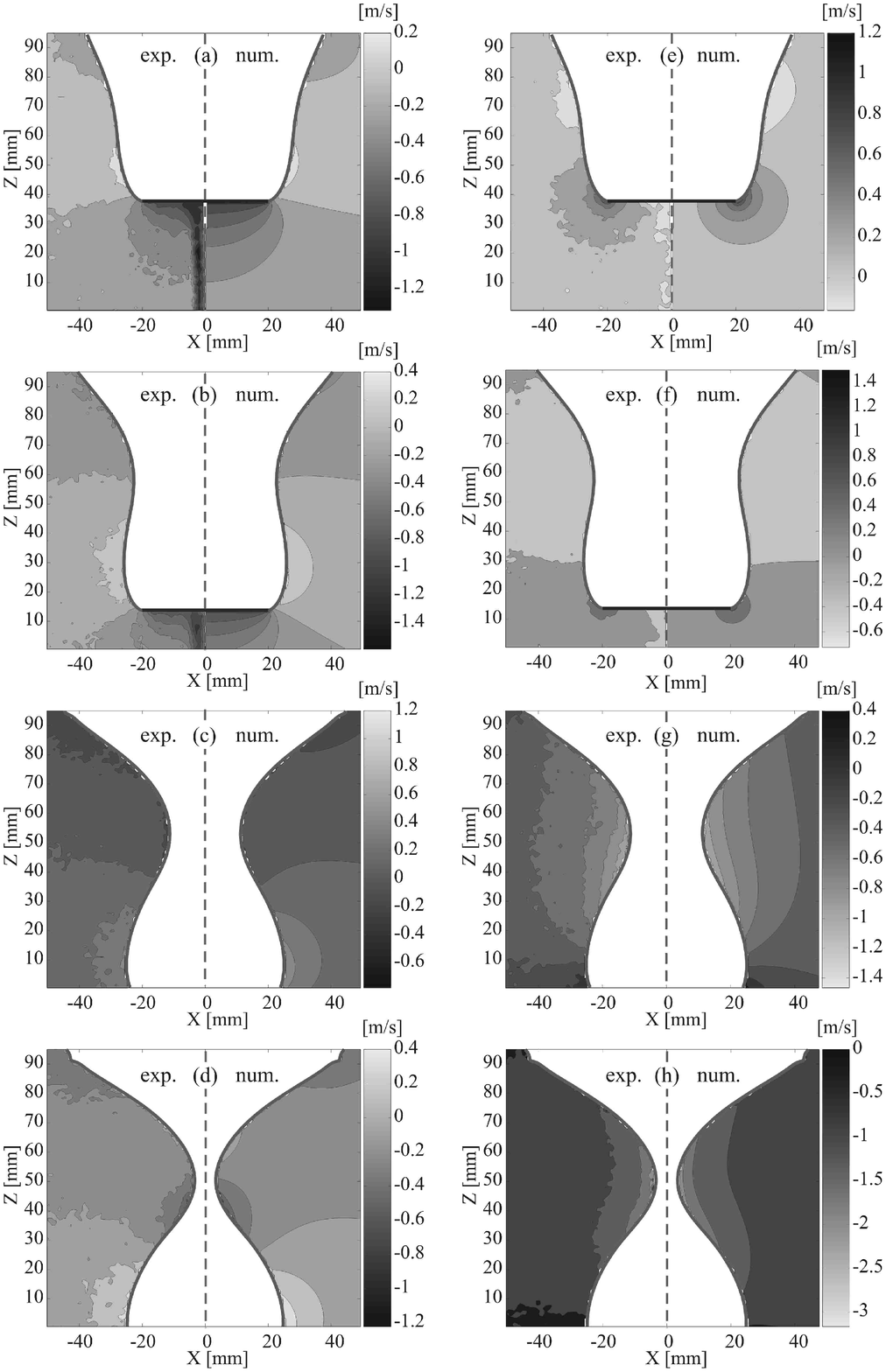}
\caption{The axial (left figures) and radial (right figures)
components of the flow field, representing the same velocity data as
in Fig.~\ref{fig:PIV global} and taken at the same times $\tau = 49$
ms (a,e), 25 ms
(b,f), 7 ms (c,g), and 1 ms (d,h). 
The four images to the left (a-d) compare a contour plot of the {\it
axial} flow component from the experiment (left side of each image)
with that of the numerics (right side of each image). The four
images on the right (e-h) show a similar comparison in a contour
plot of the {\it radial} flow component from the experiment (left
side of each image) and numerics (right side of each image). Apart
from the region where the rod is pulling down the disk in the
experiment, which is absent in the simulation, both components of
the flow field show excellent agreement between the experiments and
numerical calculations. Again this is a one-to-one comparison of
experimental and numerical data without any free parameter, neither
in velocity, space, nor time.
} \label{fig:PIV quantitative}
\end{figure}

In the previous subsection we found the experimental shape of the
impact cavity to be well described by our boundary integral
simulations if no surface seal occurs. The question we will
address in this subsection is whether the simulations also give an
accurate description of the surrounding flow field. To this end we
will measure the velocity field around the transient cavity
through high speed particle image velocimetry (PIV). These
experiments are crucial to check the validity of the boundary
integral simulations, as the presence of a solid boundary, namely
the impacting object, will induce vorticity in the flow. We will
compare the experimental flow field to the boundary integral
results and
finally investigate the radial flow at the depth of closure in more detail.\\

To perform the PIV measurements, the fluid is seeded with small
DANTEC Dynamics polyamid tracer particles of radius $25$ $\mu$m and
density $1030$ kg/m$^3$ which follow the flow. A laser sheet shines
from the side through the fluid, creating an illuminated plane
through the symmetry axis of the cavity. The light scattered by the
particles is captured by a high speed camera at a frame rate of 6000
frames per second and a resolution of 1024x512 pixels. The series of
recorded images is then correlated by multipass algorithms, using
DaVis PIV software by LaVision, in order to determine the flow field
in a plane in the liquid. The correlation was performed in two
passes at sub-pixel accuracy, using 64$\times$64 pixels and
32$\times$32 pixels interrogation windows. The windows overlap by
50\%, resulting in one velocity vector every 16$\times$16
pixels.\\

In order to obtain high resolution PIV measurements of the flow
around the cavity, we made use of the reproducibility of the
experiment. The left side of each of the images of Fig.~\ref{fig:PIV
global} shows the flow around the expanding void by combining the
results of four separate PIV measurements at different depths. In
this fashion PIV experiments were performed for a field of view of
96 mm $\times$ 56 mm at a spatial resolution of 0.9 mm (In
Fig.~\ref{fig:PIV global} only 0.7\% of the measured vector field is
shown). This high resolution makes it possible to simultaneously
compare the global flow, as well as the smaller flow structures at
the pinch--off depth and the disk's
edge.\\

\begin{figure}
\begin{center}
\includegraphics[width=0.8\textwidth]{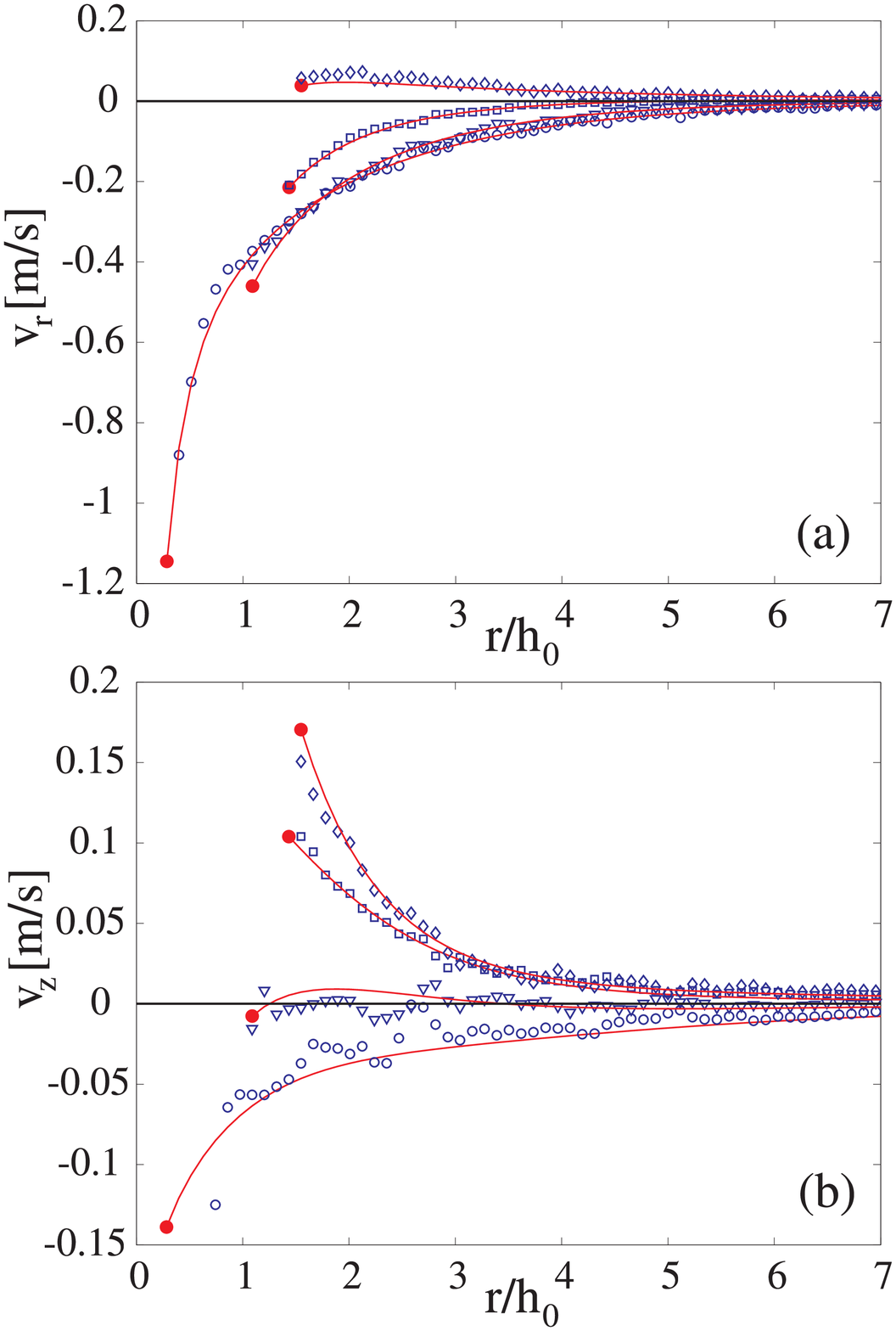}
\caption{Radial (a) and axial (b) component of the flow at the depth
of closure $z_{coll}$ for a disk of radius $h_{disk}$ = 10 mm which
impacts at a velocity of $V = 1.0$ m/s (Fr = 10.2). The symbols show
the result obtained from the PIV measurements at different times,
$\lozenge : \tau =$ 30.5 ms; $\square : \tau =$ 20.5 ms;
$\triangledown : \tau =$ 10.5 ms; and $\circ : \tau =$ 0.5 ms, and
the solid red lines are the corresponding numerical results from the
BI simulation. Note that these lines end on the cavity surface
(solid red dots). The PIV data is an average of 6 subsequent
measurements obtained from the high speed PIV recordings at 6000
frames per second. In consequence, the $v_r$ and $v_z$ velocity
components shown here are the average over one millisecond. }
\label{fig:PIV far field}
\end{center}
\end{figure}

The right side of each image of Fig.~\ref{fig:PIV global} shows
the numerically obtained cavity profile and flow field. At first
sight there appears to be a good agreement, but one would like to
obtain a more quantitative comparison between experiment and
simulation. This is provided in Fig.~\ref{fig:PIV quantitative},
which shows contour plots of the axial flow component
(Fig.~\ref{fig:PIV quantitative}a-d) and the radial flow component
(Fig.~\ref{fig:PIV quantitative}e-f) obtained from the PIV
measurements (at the left side of each image) and boundary
integral simulations (at the the right side of each image). From
this figure it is clear that the magnitude as well as the topology
of the flow are in excellent agreement. Figures~\ref{fig:PIV
global} and \ref{fig:PIV quantitative} are a one--to--one
comparison between simulation and experiment, and we
stress once more, without the use of any free parameter.\\

In addition to the above, the experimental pictures of
Fig.~\ref{fig:PIV quantitative} reveal that our initial assumption
to neglect the influence of the rod on the flow (see Section
\ref{sec:experiment}) is correct. The rod itself is clearly visible
in the experimental snapshot of Fig.~\ref{fig:PIV global}a and the
PIV software has correctly detected its downward movement, as can be
seen in Figs.~\ref{fig:PIV quantitative}a--b. From the same figures
we also conclude that outside a thin region around the rod the flow
remains unchanged. Most importantly, the outward flow at the edges
of the disk in Fig.~\ref{fig:PIV quantitative}e, which is
responsible for the expansion of the void, is unaffected by the
presence of the rod. This can be understood from the fact that below
the disk the radial flow component decays quickly towards the center
of the disk whereas the vertical component in the center is equal to
the disk speed. As the fluid in the central region hardly moves with
respect to the disk, the presence of the rod has a negligible effect
on the flow field. We conclude that the boundary layer of the rod
makes no
contribution whatsoever to the flow field around the cavity.\\

In Fig.~\ref{fig:PIV far field} the experimental flow at the
closure depth is compared with numerics up to 7 disk radii in
radial direction in order to obtain a more quantitative measure of
the magnitude of the deviations between the numerical and
experimental flow field. We find this deviation to be typically of
the order of 0.01 m/s, but it can be slightly larger if the flow
velocity is small. This larger inaccuracy at low flow velocities
is generic to the PIV method and can most clearly be seen to occur
for $\tau =10.5$ ms in Fig.~\ref{fig:PIV far field}b. Overall, a
very good agreement is found between the far field flow
in the numerics and experiments.\\

Both in the experiment and simulation we observe that during the
expansion of the void the magnitude of the outward radial flow
falls off with the distance to the symmetry axis
(Fig.~\ref{fig:PIV far field}a). However, once the cavity starts
to collapse inward there will be a region around the cavity where
the (radial) direction of the flow is reversed and there will be
an axisymmetric curved plane (manifold) where the radial flow
component vanishes. Here this happens between $\tau =30.5$ ms and
$\tau =20.5$ ms (cf.
Fig.~\ref{fig:PIV far field}a). 
In subsection \ref{sec:model_flow} we will discuss in detail how
this reversal of the radial flow expresses itself in the radial
dynamics of the cavity.


\section{Modeling the cavity dynamics} \label{sec:model2}

In this section we will first derive a simple analytical model for
the radial dynamics of the transient cavity. Secondly, we will
investigate the surrounding flow, which enters the model through
two of the free parameters and causes an asymmetry of the
collapse. In the last part of this section we compare the model to
the radial dynamics of the cavity observed in experiment and
simulation.

\subsection{A model for the radial cavity dynamics} \label{sec:model_deriv}

\begin{figure}
\begin{center}
\includegraphics[width=0.5\textwidth]{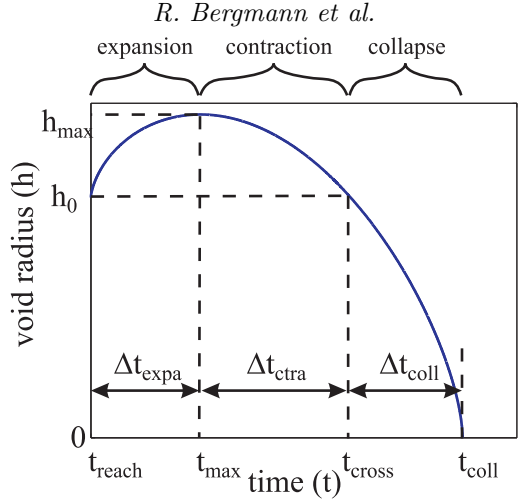}
\caption{Schematic representation of the three stages in the radial
dynamics of the cavity at a constant depth $z$. The first two stages
(corresponding to the time interval $\Delta t_{expa} + \Delta
t_{ctra}$) are governed by a forcing of the flow by hydrostatic
pressure. We can distinguish an expansion and a contraction phase
for which the model parameter $h_{\infty}$ differs considerably. In
the third stage (corresponding to $\Delta t_{coll}$) the collapsing
void accelerates towards the singularity (pinch-off) in which
inertia takes over as the driving factor.} \label{fig:cavity_theory}
\end{center}
\end{figure}

The full analytical modeling of a cylindrical symmetric collapse
of the transient cavity presents the difficulty of a coupling
between the free surface and the flow surrounding the cavity. To
tackle this difficulty we propose the convenient simplification of
dividing the problem up into a set of quasi two--dimensional
problems. If the axial component of the flow is small compared to
the horizontal flow components, we can approximate the flow as to
be confined to the horizontal plane. In this way an equation for
the collapse of a two--dimensional bubble will suffice to
describe the cavity dynamics at an arbitrary depth.\\

To derive such an equation we will closely follow a derivation given
in \cite{ogu93} and \cite{loh04a}. The argument starts by writing
the Euler equation in cylindrical coordinates, thereby neglecting
the vertical flow component and its derivatives. This means that we
assume the flow to be quasi two-dimensional at any depth along the
cavity. The azimuthal components can be ignored due to the axial
symmetry, leaving the following equation
\begin{equation}
\frac{\partial v_r}{\partial t} + v_r \frac{\partial v_r}{\partial
r} = -\frac{1}{\rho}\frac{\partial p}{\partial r}\,,
\label{eq:euleri}
\end{equation}
where $\rho$ denotes the density of the liquid. 
Under the above assumption of negligible $v_z$ and thus $\partial
v_z/\partial z$, the continuity equation and the boundary
conditions on the surface of the void lead to a second equation
\begin{equation}
 rv_r(r,t) = h(t)\dot h(t)\,.
 \label{eq:conti}
\end{equation}
Here, $h(t)$ is the radius of the cavity and its derivative $\dot
h(t)$ the velocity of the cavity wall at any depth $z$ below the
surface. 
Substituting
Eq.~(\ref{eq:conti}) into Eq.~(\ref{eq:euleri}) gives
\begin{equation}
\frac{\partial}{\partial t}\left(\frac{h\dot h}{r}\right) +
\frac{\partial}{\partial
r}\left(\frac{1}{2}v_r^2+\frac{p}{\rho}\right) = 0.
\end{equation}
We can integrate this equation over $r$ from the cavity wall $h$
to a reference point $h_{\infty}$, where the flow is taken to be
quiescent. 
This integration yields a Rayleigh-like equation for the void radius
at a fixed depth $z$,
%
\begin{equation} \label{eq_ray1a} \Bigg[ \frac{d(h\dot h)}{dt }
\Bigg]\log {h\over h_\infty} + {1\over 2} {\dot h}^2 =gz \,.
\end{equation}
Here, we have used the fact that the pressure ($p_\infty$) driving
the collapse of the cavity is provided by the hydrostatic pressure
$\rho g z$, where $z$ denotes the depth below the fluid surface.
Close to the collapse, the quantity $h_\infty$ can be interpreted as
the length scale related to the matching of an inner and outer flow
region. In the (inner) region near the neck the induced flow looks
like a collapsing cylinder as described by Eq.~(\ref{eq_ray1a}),
whereas in the (outer) region far from the neck, the flow resembles
that of a dipole (plus its image in the free surface). A complete
description of the flow would require the matching of these two
regions, where $h_\infty$ would be determined in the process as the
cross--over length scale. $h_\infty$ would thus be expected to be of
the order of a typical length scale of the process, such as the
distance of the cavity surface to the plane where the radial flow
vanishes (see Fig.~\ref{fig:PIV far field}a). Therefore, strictly
speaking, $h_\infty$ is a function of the Froude number and time. In
the model presented below we follow a different, simplified route
and set $h_\infty$ to a constant value (a time averaged value of
its dynamics).\\

We will now use Eq.~(\ref{eq_ray1a}) to analyze the radial dynamics
from the initial impact of the disk $t_0$ to the time of closure of
the cavity $t_{coll}$ at arbitrary depth $z$. In order to obtain an
analytical approximate solution, we decompose the cavity dynamics
into three different stages, depicted schematically in Figure
\ref{fig:cavity_theory}. In this figure the time intervals
corresponding to the different stages are denoted as $\Delta
t_{expa}$, $\Delta t_{ctra}$, and $\Delta t_{coll}$ respectively.
In the first two stages, during $\Delta t_{expa}$ and $\Delta
t_{ctra}$, the dynamics is dominated by the hydrostatic pressure
forcing. In these stages we observe that the water is first pushed
aside by the passing disk, creating an expanding void. At the
maximum radius $h_{max}$, the expansion is halted and the void
starts to contract. $h_{max}$ is typically of the order of $h_0$,
e.g., for Fr = 3.4 and Fr = 200 we find respectively $h_{max}\approx
1.3 h_0$ and $h_{max}\approx 2.4 h_0$. Since $\dot h(t_{max})=0$, we
can assume that $\dot h(t)$ is small during this expansion and
contraction and we can neglect the second term in
Eq.~(\ref{eq_ray1a}) leading to \beq \label{eq:hydrostatic}\Bigg[
\frac{d(h\dot h)}{dt } \Bigg]\log {h\over h_\infty}  =gz \, .\eeq
Since $\log(h/h_\infty)$ varies very slowly in the first regimes, we
equate $\log(h/h_\infty)\approx\log(h(t_{max})/h_\infty)$ and we
solve Eq.~(\ref{eq:hydrostatic}) using $h(t_{max}) = h_{max}$ and
$\dot h(t_{max}) = 0$, leading to a parabolic approximation for
$h^2$, \beq
 \label{eq:parabola}h^2(z,t) = h^2_{max} - \frac{gz}{\beta} (t
- t_{max})^2 \, ,\eeq with $\beta \equiv -\log ( h_{max}/h_\infty$).
The above equation holds for both the expansion stage, the time it
takes for the void to grow from $h_0$ to $h_{max}$, and the
contraction stage, the time it takes to shrink back to $h_0$.\\

In the third stage, during  $\Delta t_{coll}$, the collapsing void
accelerates towards the singularity and inertia takes over as the
dominant factor driving the dynamics of the cavity. This stage can
be described using a different
 approximation to Eq.~(\ref{eq_ray1a}). Near the collapse, $h$
approaches zero, $h_\infty$ is typically very large and thus the
logarithm diverges. The only way Eq.~(\ref{eq_ray1a}) can remain
valid is when the prefactor of the logarithm vanishes. This means
that \beq
 \frac{d(h\dot h)}{dt} = \frac{1}{2} \frac{d^2\left(h^2\right)}{dt^2}= 0\,. \eeq
Integration gives the power law of the two-dimensional Rayleigh
collapse (cf. \cite{ber06}) \beq h(t) = \sqrt{C}(t_{coll} - t
)^{1/2}\,. \label{eq:continuity}\eeq In
subsection~\ref{sec:model_freepar} the integration constant
$\sqrt{C}$ will be determined from continuity of $h$ and $\dot{h}$.

\subsection{The influence of the flow on $h_\infty$} \label{sec:model_flow}

As an intermezzo in the exposition of the model we now turn to an
important point, namely that there is a significantly different
quality to the flow in the expansion and the contraction stage. This
difference already became clear in our discussion of
Figure~\ref{fig:PIV far field} where we found that in the expansion
phase the outward radial flow simply decays with the radial
distance, whereas in the contraction phase the radial flow becomes
zero at some finite distance at which it changes direction. This is
due to the fact that the fluid flows outward until the cavity
reaches its maximum radius $h_{max}$, from where it will start to
move inward, creating a reversed-flow region around the cavity wall
which grows in time. Although in both stages hydrostatic pressure is
the dominant factor driving the dynamics of the cavity, there is
this dissimilarity in the surrounding flow which needs to be
incorporated into the model. To investigate the dynamics of this
dissimilarity in detail, we turn to the simulations from which we
can obtain the flow
field with an arbitrarily fine resolution.\\

\begin{figure}
\includegraphics[width=1\textwidth]{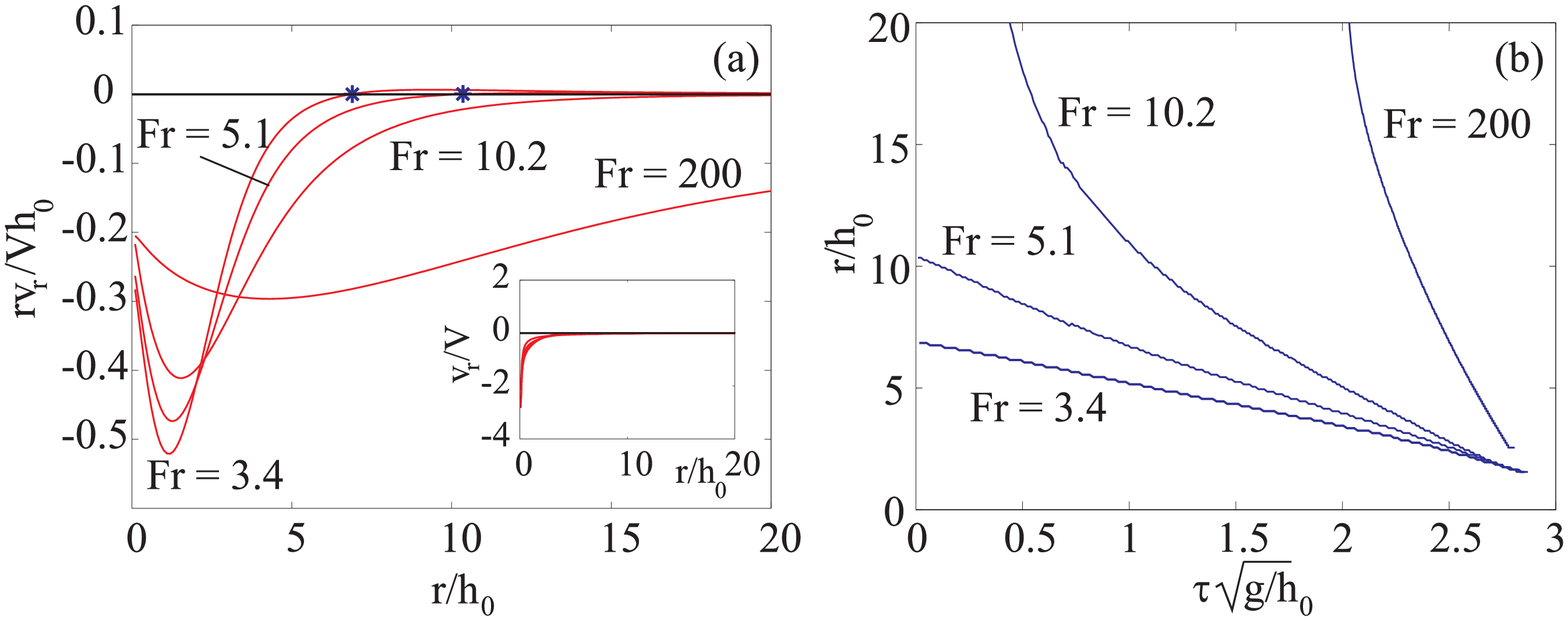}
 \caption{(a) The radial velocity component $v_r$ at the moment and depth of closure
 multiplied by the radial
 distance $r$ as a function
 of the same radial distance for different Froude numbers. The inset shows the original radial velocity data.
 The distance where the radial velocity is zero and the flow is
 stagnant in radial direction is indicated by the blue stars.
 The flow can be seen to resemble a radial sink flow more with increasing Froude number as
  the minimum of $v_rr$ decreases with increasing Froude number.\newline
  (b) The distance of the point where the radial flow reverses sign to the symmetry axis,
   determined at the
  depth of closure as a function of the normalized time remaining till
  closure $\tau=t_{coll}-t$. Below the curves the flow is directed inwards,
  above them it is directed outwards. The radial distances of the flow reversal point
  at $\tau =0$ in this figure correspond to the blue stars of figure (a). The distance of the
  point of flow reversal is 
  related to the length scales $h_\infty^{ctra}$ and $h_\infty^{expa}$, which are therefore
  expected to behave similarly in time.
} \label{fig:PIV Rvr}
\end{figure}

Figure~\ref{fig:PIV Rvr}a shows the radial flow component $v_r$
multiplied by the radial distance $r$ to the axis of symmetry at
the depth and moment of pinch-off. Since the flow at the neck
resembles locally a two dimensional sink, whose strength falls off
with $1/r$, multiplying $v_r$ with $r$ eliminates this geometrical
contribution to the flow. For the lower Froude numbers of 3.4 and
5.1 the radial flow component reverses direction at closure depth
and time at some distance $r$ (blue stars). At the higher Froude
numbers (10.2 and 200) no such point is observed within the
numerical domain, which extends to 100 disk radii in the radial
direction. This does not mean that such a flow reversal point is
absent during the complete time of the collapse, as can be seen in
Fig.~\ref{fig:PIV Rvr}b where we plot the location of the flow
reversal point as a function of time: The radial flow reversal
point comes into existence at the wall of the void at the moment
that the expanded cavity starts to collapse and the flow direction
is reversed inward. From then onwards, this point travels away
from the axis of symmetry as the collapse is approached ($\tau
\rightarrow 0$). In the same figure we also observe that for a
higher Froude number the radial flow reversal point travels
outward much faster during the cavity collapse as compared to
the low $\text{Fr}$ case.\\

\begin{figure}
 \begin{center}
\includegraphics[width=1\textwidth]{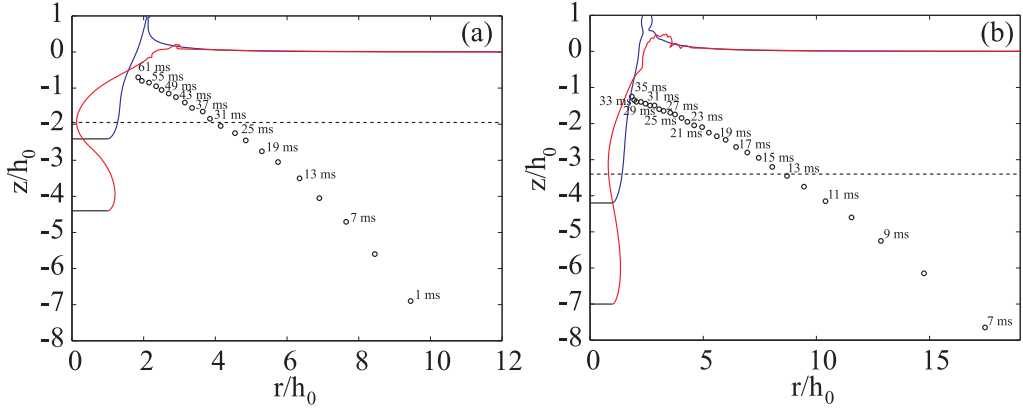}
 \caption{The open circles show the path of the stagnation point of the flow
  for Froude 3.4 (a) and 10.2 (b). For (a) the first observation of the stagnation point
  is made at $\tau =61$ ms until 1 ms before closure at intervals of 3 ms. In (b) the first observation
  is made at $\tau =35$ ms until 7 ms before closure at intervals of 1 ms. For clarity the time till closure is indicated only for every second
  observation. Further more, the void profiles at the time of
  the first (blue) and last (red) observation are shown. The
 depth of closure is indicated by the dotted line.
} \label{fig:stagnation point} \end{center}
\end{figure}

The reversed-flow region can be characterized by a stagnation point
(or saddle point) in the ($r$,$z$)-plane, corresponding to a circle
in three dimensions, where both the axial and the radial velocity
components change sign. In Fig.~\ref{fig:stagnation point} the path
of this stagnation point is shown for Fr = 3.4 (a) and Fr = 10.2
(b). For both simulations the stagnation point not only moves away
from the axis of symmetry as the pinch-off is approached, but it is
also seen to move down in the axial direction and at some point even
to cross the depth of closure. A similar path of the stagnation
point is observed for {\it all} the simulations of Fig.~\ref{fig:PIV
Rvr} and only at one instant during the collapse of the cavity does
the radial flow reversal point at closure depth truly coincide with
the stagnation
point.\\

The above leads us to three observations which are relevant for our
model of the cavity collapse: (i)~Since radial flow reversal at
closure depth occurs when the cavity starts to collapse, the
topology of the flow differs between the expansion and contraction
stage. Since $h_{\infty}$ is the radial distance at which the flow
can be assumed to be quiescent ($\mathbf{v} = 0$ and $p=p_\infty =
\rho g z$) it is related to the structure of the surrounding flow,
and it is therefore justified to assume different values of
$h_{\infty}$ in the respective stages. We will take $h_{\infty}
\equiv h_{\infty}^{expa}$ in the expansion stage and $h_{\infty}
\equiv h_{\infty}^{ctra}$ in the contraction stage of the model.
Just like $h_\infty$ in subsection~\ref{sec:model_deriv},
$h_\infty^{expa}$ and $h_\infty^{ctra}$ are set to a constant value,
representing the time averaged behavior of $h_\infty$ in each
respective stage. (ii)~As the distance of the radial flow reversal
during the contraction moves away faster at higher Froude number,
presumably a higher value for $h_{\infty}^{ctra}$ needs to be taken
for larger Froude numbers. (iii)~In \cite{ber06} we found that there
are two scaling regimes for the neck radius, a first regime where
the neck radius scales as a pure power law of time (as in
Eq.~\ref{eq:continuity}), and a second regime, where a logarithmic
correction of time has to be taken into account. The crossover
between both regimes is given by the length scale
$h_{max}^2/h_\infty^{ctra}$. As we find from Fig.~\ref{fig:PIV
Rvr}b, for all Froude numbers the distance of the radial flow
reversal increases as the pinch-off is approached. Although in
theory we assume $h_{\infty}^{ctra}$ to be constant, in reality
$h_{\infty}^{ctra}$ thus increases as the pinch--off is approached.
This means the cross--over length scale
$h_{max}^2/h_\infty^{ctra}$ decreases with time.\\
Therefore the time needed for the collapsing neck to decrease to
$h_{max}^2/h_\infty^{ctra}$ will be longer as compared to the
assumption of a constant (initial) value for $h_\infty^{ctra}$ and
may even never reach this second regime. The effect is stronger for
increasing Froude number, since the radial flow reversal point at
closure
depth moves away faster and further at higher Froude number.\\

\subsection{The free parameters of the model}\label{sec:model_freepar}

In this subsection we continue our derivation of a simplified model
for the radial cavity dynamics started in
subsection~\ref{sec:model_deriv}. As argued in the previous
subsection it is justified to assume different (constant) values for
$h_\infty$ during the expansion and contraction stage of the void.
We therefore introduce different values for $\beta$ in
Eq.~(\ref{eq:parabola}), depending on whether we are in the
expansion or in the contraction stage
\begin{eqnarray} \label{eq:beta}
\beta = \Big \{
\begin{array}{cc}
  \beta_{expa} \equiv -\log
(h_{max}(z)/h_{\infty,expa}) & t<t_{max} \\
  \beta_{ctra} \equiv -\log
(h_{max}(z)/h_{\infty,ctra}) & t>t_{max} \\
\end{array}
\, .
\end{eqnarray}
Note that with this definition $\beta_{expa}$ and $\beta_{ctra}$ are
positive quantities as for both holds
$h_{\infty,expa},h_{\infty,ctra} \gg h_{max}$. Secondly, the fact
that $\beta$ depends only logarithmically on $h_{\infty}$
furthermore justifies approximating the time-dependent quantity
$h_{\infty}(t)$ by its time-average $h_{\infty}$.

Now, to determine $h_{max}$, or rather the time it will take to get
there from the time the disk passes at $t=t_{reach}$, we need the
radial velocity of the initial expansion at $t = t_{reach}$ (see
Fig.~\ref{fig:cavity_theory}). A reasonable assumption (and similar
to the proposition of \cite{Duc06}) is that the disk displaces water
from underneath itself to the sides at a velocity directly
proportional to its downward velocity. Therefore, we have \be \dot
h(t_{reach})=\alpha_{expa} V\,.\label{eq:alpha_expa}\ee For the
velocity at the end of the contraction phase at $t_{cross} =
t_{reach} + \Delta t_{expa} + \Delta t_{ctra}$, we write in a
similar fashion \be \dot h(t_{cross})=-\alpha_{ctra}
V\,.\label{eq:alpha_ctra}\ee Clearly, both $\alpha_{expa}$ and
$\alpha_{ctra}$ are again positive quantities.

The analytical model for the radial cavity dynamics given by
Eq.~(\ref{eq:parabola}) and Eq.~(\ref{eq:continuity}) thus has
four unknown parameters $\alpha_{expa}$, $\beta_{expa}$,
$\alpha_{ctra}$, and $\beta_{ctra}$. The value of C in
Eq.~(\ref{eq:continuity}) follows from the fact that the
trajectory and its derivative in Fig.~\ref{fig:cavity_theory}
should be continuous. We assume the collapse regime to start at
the end of the contraction phase, where we have $h(t_{cross})=h_0$
and $\dot h(t_{cross}) = -\alpha_{ctra} V$. From these conditions,
the value of $C$ is readily obtained, \beq C = 2h_0\alpha_{ctra}
V\,. \eeq However, for given $\alpha_{expa}$, $\beta_{expa}$, and
$\beta_{ctra}$, the constant $\alpha_{ctra}$ is also uniquely
determined by the continuity of the trajectory and its derivative
at $h(t_{max}) = h_{max}$ , which gives
\be\alpha_{ctra}=\alpha_{expa}\sqrt{\beta_{expa}/\beta_{ctra}}\,,\label{eq:alpha_relation}\ee
and leaves only $\alpha_{expa}$,
$\beta_{expa}$, and $\beta_{ctra}$ to be determined.\\

Summarizing, the time evolution of the cavity at depth $z$ is
described by the following three equations
\begin{eqnarray}
 h(z,t) =& \sqrt{h^2_{max} - \frac{gz}{\beta_{expa}} (t - t_{max})^2} & \qquad \text{for  } t_{reach} < t \leq t_{max} \,,\label{eq:traj1}\\
 h(z,t) =& \sqrt{h^2_{max} - \frac{gz}{\beta_{ctra}} (t - t_{max})^2} & \qquad \text{for  }t_{max} < t \leq t_{cross} \,,\label{eq:traj2}\\
 h(z,t) =& \sqrt{2h_0\alpha_{ctra}V}\sqrt{t_{coll} - t} & \qquad \text{for  } t_{cross} < t \leq t_{coll}\,, \label{eq:traj3}
\end{eqnarray}
where the times $t_{reach}$, $t_{max}$, $t_{cross}$, and $t_{coll}$
are readily related to the impact time $t = 0$ (which will be done
explicitly in section~\ref{sec:characteristics}) and $h_{max}$ is
given by \beq h_{max}(z) = h_0 \sqrt{1 +
\alpha_{expa}^2\beta_{expa}\frac{V^2}{gz}}\,, \label{eq:hmax}\eeq as
can be easily derived, e.g., from Eq.~\ref{eq:traj1} together with
its boundary conditions $h(z,t_{reach}) = h_0$ and
$\dot{h}(z,t_{reach}) = \alpha_{expa}V$.

\subsection{Validation of the model}\label{sec:modelvalidation}

We will now compare the dynamics of the radius of the void at
closure depth with the theoretical prediction of
Eq.~(\ref{eq:parabola}) and Eq.~(\ref{eq:continuity}) to validate the model and quantify
the influence of the flow reversal on
$\beta_{expa}$ and $\beta_{ctra}$.\\

The parameter $\alpha_{ctra}$ is eliminated by the relation
Eq.~(\ref{eq:alpha_relation}), leaving three parameters to match
Eqs.~(\ref{eq:traj1})--(\ref{eq:traj3}) to the radial dynamics from
the boundary integral simulations. 
At first sight, one could assume that the initial outward velocity
$\alpha_{expa} V$ could be easily obtained from simulation or
experiment, since it is observed as the angle at which the free
surface leaves the disk. However, when this angle is investigated in
closer detail, it is found to strongly depend on the distance from
the disk's edge over which it is measured. In the numerics, close
enough to the disk's edge, the free surface even becomes nearly
parallel with the disk. This means that although $\alpha_{expa} V$
is useful as a (theoretical) concept, it is not directly measurable
and should therefore be determined by a fitting routine.\\ 

\begin{figure}
 \begin{center}
\includegraphics[width=1\textwidth]{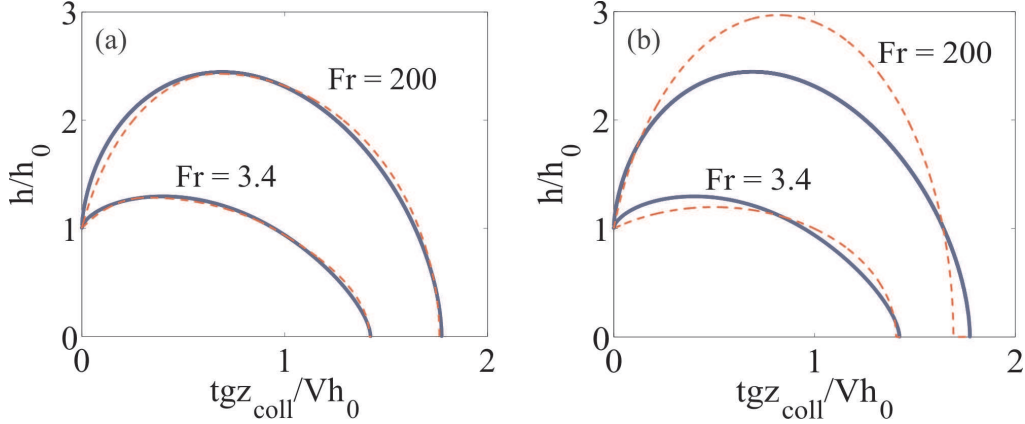}
 \caption{(a) The time evolution of the radius of the cavity at closure depth for different Froude numbers.
 The solid blue lines represent the simulation results and the dashed red lines correspond to a least square
 fit of the approximation given by Eqs.~(\ref{eq:traj1})-(\ref{eq:traj3}). 
(b) The same numerical time evolution data as in (a) (solid blue
lines) are now approximated by the model proposed by \cite{Duc06},
which consists of Eqs.~(\ref{eq:traj1})--(\ref{eq:traj2}) with
$\beta_{expa}$ and $\beta_{ctra}$ set to 1, again from a least
squares fit (dashed red lines). In both (a) and (b) time has been
rescaled by a factor $Vh_0/gz_{coll}$ in order to show the results
for the two Froude numbers in a single plot.
} \label{fig:R T}
\end{center}
\end{figure}

The three parameters $\alpha_{expa}$, $\beta_{expa}$, and
$\beta_{ctra}$ are determined by a least square fit to the radial
dynamics of the cavity at closure depth obtained from the
simulations. Fig.~\ref{fig:R T}a shows the comparison between
these fits of Eqs.~(\ref{eq:traj1})--(\ref{eq:traj3}) (red dashed
line) and the simulations (blue solid line) at two different
Froude numbers of 3.4 and 200. The approximation is found to be in
excellent agreement throughout the collapse, faithfully
reproducing the maximum expansion of the cavity and the complete
time of collapse.

\begin{figure}
\begin{center}
\includegraphics[width=1\textwidth]{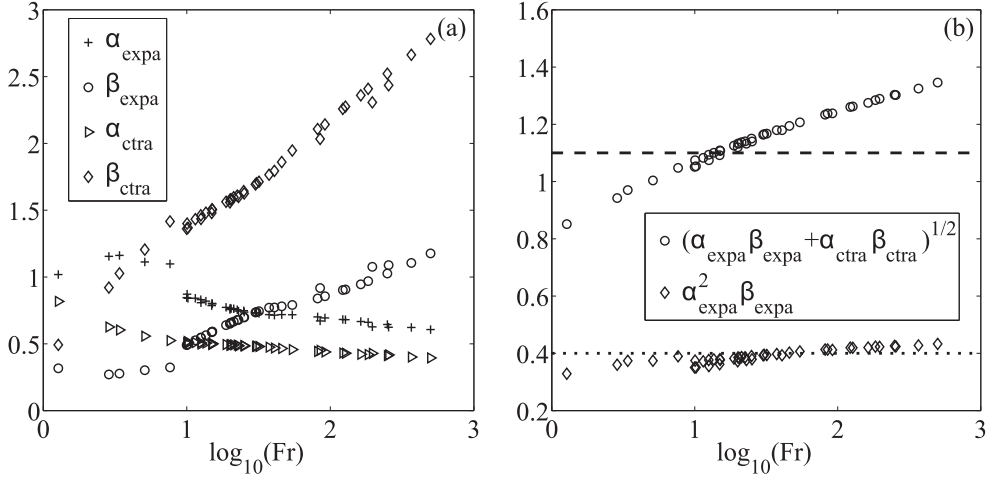}
\caption{(a) The parameters $\alpha_{expa}$, $\beta_{expa}$,
$\alpha_{ctra}$, and $\beta_{ctra}$ obtained from the fitting
routine used in Fig.~\ref{fig:R T}a. All are found to weakly depend
on the Froude number (note the logarithmic scale for $\text{Fr}$).
(b) The quantities $\sqrt{\alpha_{expa}\beta_{expa} +
\alpha_{ctra}\beta_{ctra}}$ and $\alpha_{expa}^2\beta_{expa}$
calculated from (a). Whereas $\sqrt{\alpha_{expa} \beta_{expa} +
\alpha_{ctra}\beta_{ctra}}$ is found to weakly depend on the Froude
number, $\alpha_{expa}^2\beta_{expa}$ has a nearly constant value of
0.40. (horizontal dotted line). In comparison with
$\sqrt{\alpha_{expa}\beta_{expa} + \alpha_{ctra} \beta_{ctra}}$, the
horizontal dashed line indicates the value $C_1 = 1.10$ obtained
from a best fit to the closure depth data of
Fig.~\ref{fig:closure_depth}.} \label{fig:sqrt}
\end{center}
\end{figure}

In Figure~\ref{fig:sqrt}a we find the parameters $\alpha_{expa}$,
$\beta_{expa}$, and $\beta_{ctra}$ as a function of the Froude
number, determined by repeating the fitting routine described
above for many impact velocities. All are found to weakly depend
on the Froude number (note that a logarithmic scale has been used
for $\text{Fr}$). For completeness we also plot the derived
quantity $\alpha_{ctra}$, calculated from Eq.~\ref{eq:alpha_relation}.\\

If in Eqs.~(\ref{eq:traj1})--(\ref{eq:traj2}) the constants
$\beta_{expa}$ and $\beta_{ctra}$ are set to 1 and therefore by
Eq.~(\ref{eq:alpha_relation}), $\alpha_{ctra}=\alpha_{expa}$, we
arrive at the cavity dynamics proposed by \cite{Duc06} for
impacting spheres and cylinders. These dynamics are shown in
Fig.~\ref{fig:R T}b with the only free parameter $\alpha_{expa}$
also determined by a least square fit to the data. This
approximation is seen to qualitatively reproduce the trend for the
maximum expansion and collapse time, but fails to capture the
exact values. It is also conceptually different, as \cite{Duc06}
propose the cavity dynamics to be symmetric around the maximum
expansion, while our model captures the asymmetry around this
point in time that is also found in experiment and simulation. Our
solution Eqs.~(\ref{eq:traj1})--(\ref{eq:traj2}) is explicitly not
symmetric, since it allows for
different values of $\beta$ at $t<t_{max}$ and $t>t_{max}$.\\

To conclude this section we return to the first two observations we
made at the end of subsection~\ref{sec:model_flow} on the motion of
the stagnation point and the plausible consequences for $h_\infty$.
(i) The flow reversal which occurs when the cavity starts to
collapse indeed introduces an asymmetry in the behavior around the
maximum expansion. This is clearly observed in the radial dynamics
of Fig.~\ref{fig:R T}, especially for $\text{Fr} = 3.4$. (ii) As the
distance of the radial flow reversal during the collapse moves away
faster at higher Froude number (see Fig.~\ref{fig:PIV Rvr}b), we
indeed have to introduce a larger $h_{\infty}^{ctra}$ (corresponding
to a larger $\beta_{ctra}$) for higher Froude number in the fit of
Fig.~\ref{fig:R T}a to account for this effect.\\

\section{Characteristics of the Transient
Cavity}\label{sec:characteristics}

Now that we derived a simplified model for the radial dynamics of
the  cavity, we will use it, together with the simulations and
experiments, to investigate the following key characteristics of the
transient cavity: (i) the depth of the pinch-off and the depth of
the disk at the moment of pinch--off (subsection \ref{sec:closure
depth}), and (ii) the amount of air entrained through the cavity
collapse (subsection \ref{sec:air entrainment}). Besides this we
elaborate on our previous findings from \cite{ber06} in the
Appendix~\ref{appendixA}. There we revisit the dynamics of the
cavity at closure depth (Appendix \ref{sec:rad_closure}) and the
cavity shape around the minimal radius (Appendix \ref{sec:neck
radius}). Finally, in Appendix~\ref{app:minimal radius} we discuss
the time evolution of the vertical position of the minimal cavity
radius and place it within the context of the model.

\subsection{Closure depth} \label{sec:closure depth}

\begin{figure}
\begin{center}
\includegraphics[width=0.65\textwidth]{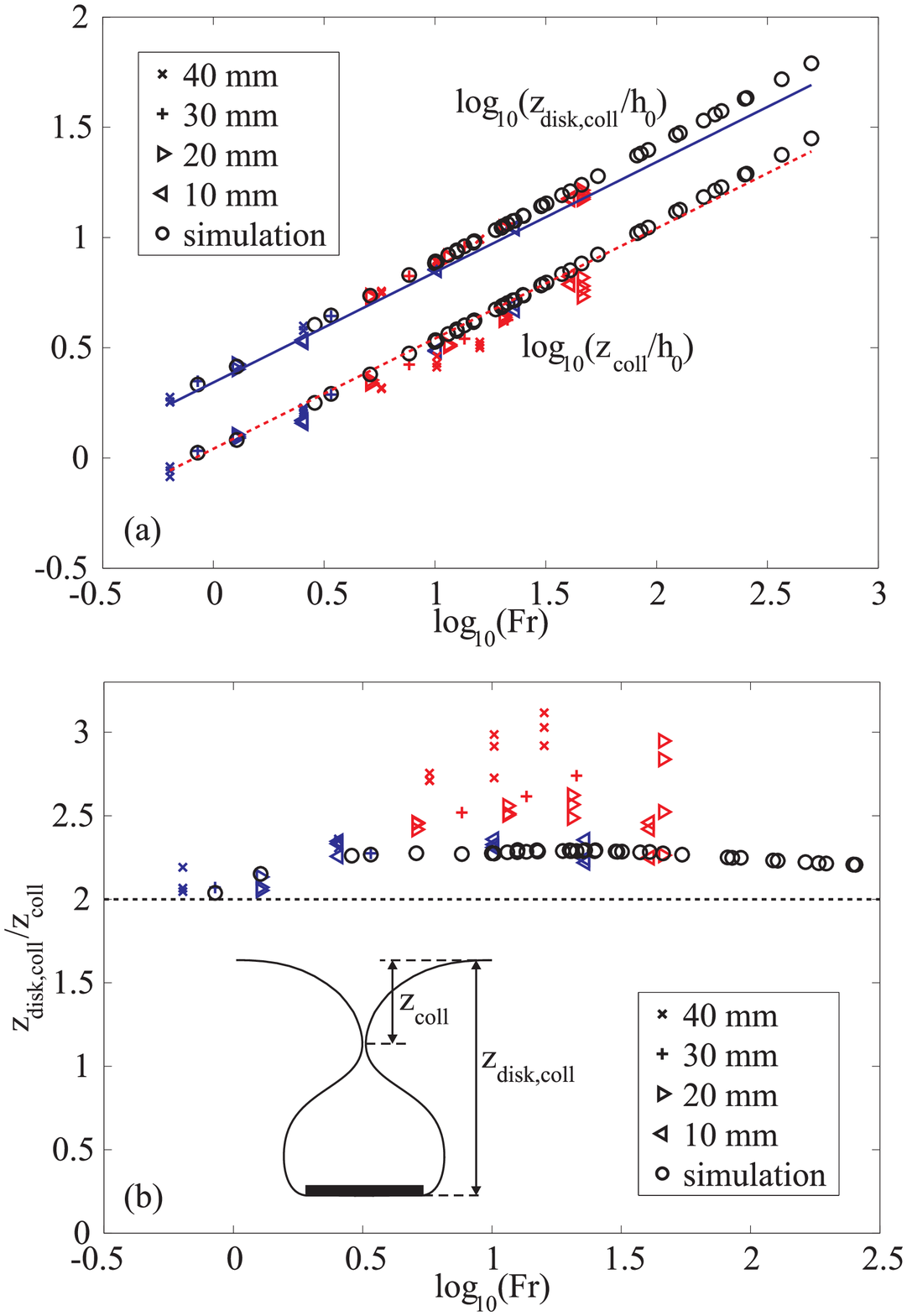}
\caption{(a) Double logarithmic plot of the depth at which the void
collapses $z_{coll}$ and the depth $z_{disk,\,coll}$ of the disk at
collapse time for experiments with four different disk radii (see
legend) and for the boundary integral simulations (open circles),
all as a function of the Froude number. Experiments in which a
surface seal occurs during the collapse are indicated by red
symbols, the experiments without a surface seal by blue symbols.
Here, a surface seal is said to occur if at some point in time due
to air suction the splash closes onto itself and the cavity is
completely sealed off. The experiments without a surface seal are
found to agree well with the numerically obtained values (open
circles) and the theoretical prediction for the scaling of
$z_{coll}/h_0 = C_1 Fr^{1/2}$ (red dashed line) and
$z_{disk,\,coll}/h_0 = 2 C_1 Fr^{1/2}$ (blue solid line), with $C_1=
1.10$ obtained from a fit to the data of $z_{coll}$. The experiments
for which a surface seal occurs are seen to slightly deviate from
this prediction.
\newline
(b) The ratio of the depth of the disk at the time of pinch-off
$z_{disk,\,coll}$ and the pinch-off depth $z_{coll}$ for different
disk radii as a function of the Froude number. The experiments
without a surface seal (blue symbols) agree well with the numerical
results (open circles). The ratio for the numerical result and
experiments without a surface seal lie close to the predicted value
of 2 indicated by the black dashed line. The experiments in which a
surface seal occurs are again indicated by the red symbols and found
to deviate more with increasing Froude number for a fixed disk size.
The inset shows the definition of the depths $z_{coll}$ and
$z_{disk,coll}$ at the closure time.
} \label{fig:closure_depth}
\end{center}
\end{figure}

Following \cite{gla96}, \cite{gau98}, and \cite{Duc06} we will
characterize the shape of the cavity at pinch--off by the depth of
closure $z_{coll}$, i.e., the depth at which the pinch--off takes
place. To capture more information on the full shape of the void,
we will also investigate how $z_{coll}$ relates to the total depth
of the cavity $z_{disk}(t_{coll})=z_{disk,\,coll}$
at the time of collapse (or closure) (see the inset of Fig.~\ref{fig:closure_depth}).\\

A comprehensive argument for the scaling of $z_{coll}$ can be
obtained by following a similar procedure to the one outlined in
\cite{loh04a} for the determination of the closure depth after the
impact of a steel ball on soft sand. The difference is that whereas
in sand one can assume that due to the compressibility of the
material there is hardly any outwards motion of the sand, here we
are dealing with an incompressible fluid and the outward
expansion of the cavity needs to be taken into account.\\

The time interval between impact of the disk and collapse of the
cavity $\Delta t = t_{coll}-t_0$ at any depth $z$ consists of two
main parts: First, the disk needs an amount of time $\Delta
t_{reach}$ to reach the depth $z$. Second, just after the disk
passes there is the time $\Delta t_{void}$ it takes for the void to
form, expand, and collapse \beq \Delta t = \Delta t_{reach} + \Delta
t_{void} \,. \eeq The first term equals $\Delta t_{reach} = z/ V$
since the velocity of the disk is constant in the experiment and
simulation. In Section~\ref{sec:model2} $\Delta t_{void}$ was
decomposed into three stages as is schematically depicted in
Fig.~\ref{fig:cavity_theory}.
The collapse time can thus be written as, \begin{eqnarray} \Delta t
= \Delta t_{reach} + \underbrace{\Delta t_{expa} + \Delta t_{ctra} +
\Delta t_{coll}}_{\Delta t_{void}}\,.\end{eqnarray} To estimate
these last three timescales at arbitrary depth $z$, we turn to our
model for the cavity dynamics Eqs.~(\ref{eq:traj1})-(\ref{eq:traj3}).\\

If we combine the conditions Eq.~(\ref{eq:alpha_expa}) and
Eq.~(\ref{eq:alpha_ctra}) with the time derivative of
Eqs.~(\ref{eq:traj1}) and (\ref{eq:traj2}), we readily obtain
\begin{eqnarray} \label{eq:time_exp&ctra}
\Delta t_{expa} =\alpha_{expa} \beta_{expa} \frac{h_0V}{gz}\,,\nonumber\\
\Delta t_{ctra} = \alpha_{ctra} \beta_{ctra} \frac{h_0V}{gz}\,.\\
\end{eqnarray}
Recall that $\alpha_{ctra} =
\alpha_{expa}(\beta_{expa}/\beta_{ctra})^{1/2}$. The radial
collapse during $\Delta t_{coll}$ is in turn described by the
approximation of Eq.~(\ref{eq:traj3}). Since $h(t_{cross}) = h_0$
we find for this time interval \beq \Delta t_{coll} =
t_{coll}-t_{cross} = \frac{1}{2\alpha_{ctra}}\frac{h_0}{V}\,.\eeq
Collecting all the above time intervals, within the model the
total amount of time that passed from the impact of the disk until
the collapse of the cavity at depth $z$ is given by \beq
\label{eq:closure time} \Delta t &=& \Delta t_{reach} + \Delta
t_{expa} + \Delta t_{ctra} + \Delta t_{coll} \nonumber\\&=&
\frac{z}{V}+ (\alpha_{expa} \beta_{expa} + \alpha_{ctra}
\beta_{ctra})\frac{h_0V}{gz}+\frac{1}{2\alpha_{ctra}}\frac{h_0}{V}\,.\eeq
Now, to find the closure depth $z_{coll}$, we need to determine at
what depth the collapse will occur first, which we can do by
solving \beq \frac{d\Delta t}{dz} = 0\,.\eeq This gives \beq
\label{eq:closure depth} \frac{z_{coll}}{h_0} =
\sqrt{\alpha_{expa} \beta_{expa} + \alpha_{ctra}
\beta_{ctra}}\text{Fr}^{1/2}\,.\eeq In addition, the total depth
of the disk at the time of collapse,
$z_{disk}(t_{coll})=z_{disk,\,coll}$, can be obtained by inserting
Eq.~(\ref{eq:closure depth}) into Eq.~(\ref{eq:closure time}) to
give $z_{disk,\,coll} = V\Delta t$, or \beq \label{eq:total depth}
\frac{z_{disk,\,coll}}{h_0} = 2\sqrt{\alpha_{expa} \beta_{expa} +
\alpha_{ctra} \beta_{ctra}}\text{Fr}^{1/2}
+\frac{1}{2\alpha_{ctra}}\,. \eeq

When we compare these expressions with the experiments without a
surface seal (blue symbols) and the numerical calculations (black
circles) in Fig.~\ref{fig:closure_depth}a we find a very good
agreement with the prediction of Eq.~(\ref{eq:closure depth}). A fit
to the data of $z_{coll}$ gives $z_{coll}/h_0=C_1\text{Fr}^{1/2}$,
with $C_1=1.10$. The agreement of the experiments in which a surface
seal occurs (red symbols) deteriorates for a fixed disk size with
increasing Froude number,
since the surface seal becomes more disruptive at higher impact velocities.\\
In the same figure we find the experimental and numerical results
for the total depth of the disk at closure $z_{disk,\,coll}$. From
the apparent power-law scaling it is clear that the constant
$1/(2\alpha_{ctra})$ in Eq.~(\ref{eq:total depth}) has no
significant contribution. The total depth of the void is found to
scale as $z_{disk,\,coll}/h_0=C_2 \text{Fr}^{1/2}$, with $C_2=2.49$
close to the expected value of $C_2 =2C_1 = 2.2$ that follows from
Eqs. (\ref{eq:closure depth}) and (\ref{eq:total depth}). The fact
that the closure depth and the total depth have the same power-law
scaling $\text{Fr}^{1/2}$ indicates that the time from the initial
impact of the disk to the time of closure of the cavity does not
depend on the velocity of the impact\footnote{Within the model,
keeping the constant term in Eq.~(\ref{eq:total depth}) is
equivalent to keeping the last term in Eq.~(\ref{eq:closure time})
 which would add a $1/V$-dependence to the closure time, vanishing
 for high Froude number.}, since $\Delta
t=z_{disk,\,coll}/V=C_2\sqrt{h_0/g}$.\\

This is in agreement with the findings of \cite{gla96}, who
experimentally observed a similar scaling for the impact of a disk
on a water surface, although with a slightly different prefactor
of $C_2 \approx 2.3$. \cite{Duc06} also found the scaling of
$\Delta t$ ($\propto\sqrt{h_0/g}$) for impacting spheres and
furthermore reported $z_{disk,\,coll}/h_0=2z_{coll}/h_0\propto
\text{Fr}^{1/2}$
in agreement with our observations. \\

To investigate the data of Fig.~\ref{fig:closure_depth}a more
closely it is convenient to take the ratio of
$z_{coll}/z_{disk,\,coll}$ (see Fig.~\ref{fig:closure_depth}b).
According to Eq.~(\ref{eq:closure depth}) and Eq.~(\ref{eq:total
depth}), this ratio should scale as \beq \label{eq:ratio depths}
z_{coll}/z_{disk,\,coll} =
2+\frac{1}{2\alpha_{ctra}\sqrt{\alpha_{expa} \beta_{expa} +
\alpha_{ctra} \beta_{ctra}}}\text{Fr}^{-1/2}\approx 2\,\eeq in the
limit of large Froude number. In Fig.~\ref{fig:closure_depth}b the
ratio of $z_{coll}/z_{disk,\,coll}$ in the experiments without a
surface seal (blue symbols) and the numerical calculations (black
circles) are indeed close to the constant value of 2 (dashed black
line), but at lower Froude number it decreases slightly contrary
to the proposed scaling by the second term in Eq.~(\ref{eq:ratio
depths}). Although the second term of the ratio of
Eq.~(\ref{eq:ratio depths}) should become relevant when the Froude
number is considerably small, this is not observed in
Fig.~\ref{fig:closure_depth}b. This can be understood by noting
that in the limit of small Froude number our assumption of
non-interacting fluid layers from Eq.~(\ref{eq_ray1a}) breaks down
as gravity becomes more important and thus Eq.~(\ref{eq:ratio
depths}) is no longer valid. In Fig.~\ref{fig:closure_depth}b it
is again illustrated that the experiments with a surface seal (red
symbols) deviate more and more from the simulations without air as
the Froude
number increases.\\

The fits to the trajectories discussed in
subsection~\ref{sec:modelvalidation} provide us with the
parameters $\alpha_{expa}$, $\beta_{expa}$, and $\beta_{ctra}$
(recall that $\alpha_{ctra}$ is given by Eq.
(\ref{eq:alpha_relation})), and therewith with an
\emph{independent} way of determining the proportionality constant
$\sqrt{\alpha_{expa}\beta_{expa}+ \alpha_{ctra}\beta_{ctra}}$ of
Eq.~\ref{eq:closure depth}. Repeating this fitting procedure for
many impact velocities results in Fig.~\ref{fig:sqrt}b, where
$\sqrt{\alpha_{expa}\beta_{expa} + \alpha_{ctra}\beta_{ctra}}$ is
plotted as a function of $\log_{10}\text{Fr}$. A weak
(logarithmic) dependence on the Froude number is revealed. It can
also be seen that the value $C_1 \approx 1.10$ of the
proportionality constant in Eq.~(\ref{eq:closure depth}) found
from the fit to the closure depth data in
Fig.~\ref{fig:closure_depth} is consistent with the data when one
wants to disregard the $\text{Fr}$ dependence.

\subsection{Air Entrainment} \label{sec:air entrainment}

\begin{figure}
\begin{center}
\includegraphics[width=0.7\textwidth]{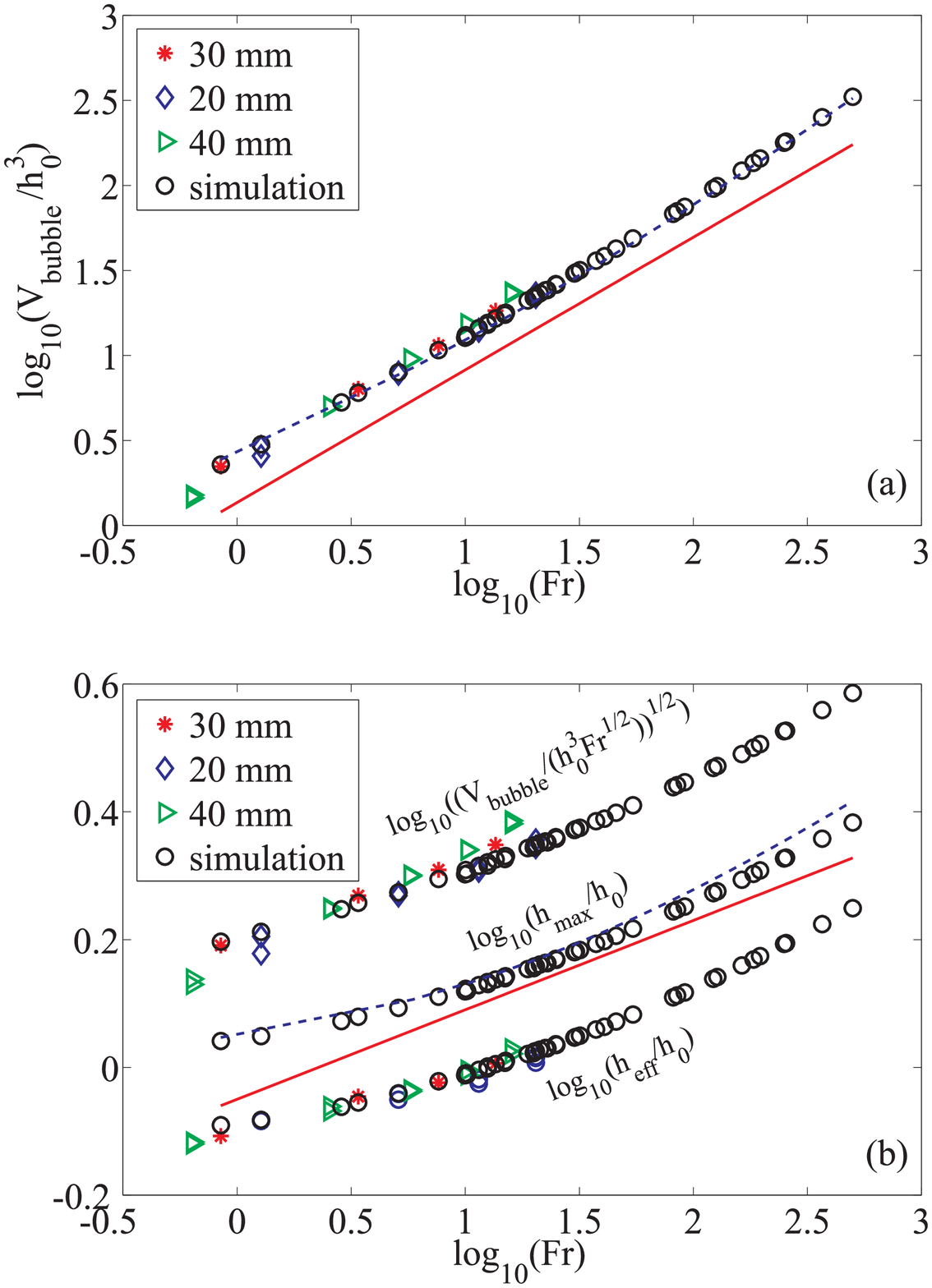}
\caption{(a) The volume of the bubble $V_{bubble}$ entrained during
the collapse of the cavity from experiments (colored symbols) and
simulations (black open circles), normalized by the cubed disk
radius $h_0^3$, as a function of the Froude number in a double
logarithmic plot. The data suggest a power-law scaling
$V_{bubble}/h_0^3 \propto \text{Fr} ^{\lambda}$ where a linear best
fit through the data between $\text{Fr} = 2.5$ and $250$ gives
$\lambda \approx 0.78$ (solid red line, shifted for clarity). The
dashed blue line corresponds to the model prediction
Eq.~(\ref{eq:bubmodscaling}).\newline (b) Double logarithmic plot of
three quantities that measure the radial length scale of the
entrapped air bubble. From bottom to top: the effective (or average)
radius $h_{\textit{eff}}$ of the bubble at pinch--off
(Eq.~(\ref{eq:effradius})), the maximum radius of the bubble
$h_{max,\,coll}$ at pinch--off, and the square root of the bubble
volume compensated for the expected scaling of its vertical
extension $[V_{bubble}/h_0\text{Fr}^{1/2}]^{1/2}$, all compensated
with the disk radius $h_0$. The dashed blue line is the model
prediction Eq.~(\ref{eq:radmodscaling}) and the solid red line
represents a power law with the scaling exponent $\lambda = 0.14$
expected from Eq.~(\ref{eq:bubble_scaling}).} \label{fig:volume
bubble}
\end{center}
\end{figure}

After pinch-off, an air bubble is entrapped, as is clearly visible
in Figs.~\ref{fig:experimental_setup}b and \ref{fig:comparison}.
The (rescaled) volume of this bubble $V_{bubble}/h_0^3$ is not
only found to solely depend on the Froude number, but also to
exhibit close to power-law scaling behavior: The scaling law for
the volume of the bubble observed in experiment and simulation is
found to be $V_{bubble}/h_0^3 \propto \text{Fr}^\lambda$, with
$\lambda =$ 0.78
(see Fig.~\ref{fig:volume bubble}a).\\
This is surprising, since for the impact of a liquid mass on a free
surface the volume of air entrained in the process scales with a
different exponent $V_{bubble}\propto \text{Fr}^{1.0}$
(\cite{pro97}). In this section we will try to shed light onto the
origin of this scaling behavior using our findings
of Section~\ref{sec:closure depth}.\\

In Section~\ref{sec:closure depth} it was found that the axial
length of the enclosed bubble at pinch--off scales roughly as
$(z_{coll}- z_{disk,\,coll})/h_0\approx 1.10 \, \text{Fr}^{1/2}$, if
we ignore the weak Froude number dependence of the prefactor
$\sqrt{\alpha_{expa}\beta_{expa} + \alpha_{ctra}\beta_{ctra}}$ (see
Fig.~\ref{fig:sqrt}b). Therefore, the scaling of the axial length
$z_{coll}- z_{disk,\,coll}$ of the enclosed bubble cannot by itself
account for the observed scaling of $V_{bubble}$. The radial length
scale $h_{rad}$ of the bubble must therefore be Froude number
dependent and should scale as
\begin{equation} \label{eq:bubble_scaling}
\frac{h_{rad}}{h_0} \propto
\left[\frac{V_{bubble}}{h_0^2(z_{disk,coll}-z_{coll})}\right]^{1/2}
\propto \left[\frac{\text{Fr}^{0.78}}{\text{Fr}^{0.50}}\right]^{1/2}
= \text{Fr}^{0.14}\,.
\end{equation}

Now what would we expect based on our simplified model? The
maximum radial expansion of the cavity at any depth $z$ is given
by $h_{max}(z)$, see Eq.~(\ref{eq:hmax}). As the depth $z_{max}$
at which the radial size of the bubble is maximal is somewhere
between the closure depth $z_{coll}$ and the depth of the disk at
closure $z_{disk,\,coll}$, we have $z_{max} \approx (z_{coll} +
z_{disk,\,coll})/2 = (3/2)\sqrt{\alpha_{expa}\beta_{expa} +
\alpha_{ctra}\beta_{ctra}}\text{Fr}^{1/2}$. If we insert this
depth into $h_{max}(z)$, Eq.~(\ref{eq:hmax}), we find
\begin{eqnarray}\label{eq:radmodscaling}
h_{rad} &\propto& h_{max}(z_{max}) \approx h_0 \sqrt{1 +
\frac{2\alpha_{expa}^2\beta_{expa}}{3\sqrt{\alpha_{expa}\beta_{expa}
+ \alpha_{ctra}\beta_{ctra}}}\text{Fr}^{1/2}} \nonumber\\
&\approx& h_0 \sqrt{1 + 0.26 \,\text{Fr}^{1/2}} \, .
\end{eqnarray}
In the last (approximate) equation we have used that
$\sqrt{\alpha_{expa}\beta_{expa} + \alpha_{ctra}\beta_{ctra}}
\approx 1.10$ and $\alpha_{expa}^2\beta_{expa} \approx 0.40$ (cf.
Fig.~\ref{fig:sqrt}b). If a power-law fit $h_{rad}/h_0$ vs.
$\text{Fr}$ is enforced on this dependence in the regime $2.5 <
\text{Fr} < 250$ one obtains the observed effective exponent $0.14$,
$h_{rad}/h_0 \propto \text{Fr}^{0.14}$. Alternatively, by taking the
square of Eq.~(\ref{eq:radmodscaling}) and multiplying with the
vertical extension $(z_{coll}- z_{disk,\,coll})$ of the bubble we
find the following prediction for the bubble volume
\begin{equation}\label{eq:bubmodscaling}
V_{bubble} \propto h_0^3\left(1 + 0.26 \,\text{Fr}^{1/2}\right)
\text{Fr}^{1/2} \, .
\end{equation}
Clearly, the model predicts power-law scaling only in the limit of
large Froude numbers. Moreover, as in this limit $V_{bubble} \propto
\text{Fr}$, the scaling prediction is in agreement with the
 \cite{pro97} result. Again, in the regime $2.5 <
\text{Fr} < 250$ the effective exponent is $0.78$. \\

We test the above prediction by looking at three different
quantities that capture the radial expansion of the cavity in
experiment and numerics. The first is the effective, or average,
radius $h_{\textit{eff}}$ of the bubble which is computed directly
from the experimental and numerical cavity profiles (i.e., without
any scaling assumption of the axial length scale) at the pinch-off
time by
\begin{equation} \label{eq:effradius}
h_{\textit{eff}}^2 =\frac{1}{\left(z_{disk,\,coll}-z_{coll}\right)}
\int_{z_{disk,\,coll}}^{z_{coll}}h^2(z)dz \, .
\end{equation}
The second quantity we look at is the maximal radius of the bubble
$h_{max,\,coll}$ at the time of pinch--off which is a more direct
measure of the expansion of the cavity. $h_{max,\,coll}$ can be
directly observed from the cavity profile at the time of pinch--off
as the maximal radius for a depth between
$z_{disk,\,coll}$ and $z_{coll}$.\\

In Figure~\ref{fig:volume bubble}b we compare these two quantities
$h_{\textit{eff}}/h_0$ and $h_{max,\,coll}/h_0$ with a third, namely
the measured $V_{bubble}$ compensated for the expected scaling
behavior of its axial extension $z_{disk,coll}-z_{coll} \propto h_0
\text{Fr}^{1/2}$, i.e., $V_{bubble}/h_0^3\text{Fr}^{1/2}$.
All of these three quantities follow the same trend, which is 
well described by the prediction Eq.~(\ref{eq:radmodscaling}) from
the model (the blue dashed line in Fig.~\ref{fig:volume bubble}b),
and close to the expected $\text{Fr}^{0.14}$ scaling which is
indicated by the solid red line. Finally, comparing the measured
bubble volume $V_{bubble}/h_0^3$ with the model prediction
Eq.~(\ref{eq:bubmodscaling}) in Fig.~\ref{fig:volume bubble}a
(dashed blue line), we find excellent agreement.\\

\section{Conclusions} \label{sec:conclusion}

In this article we investigate the purely gravitationally induced
collapse of a surface cavity created by the controlled impact of a
disk on a water surface. We find excellent agreement between
experiments and boundary integral simulations for the dynamics of
the interface, as well as for the flow surrounding the cavity. The
topology and the magnitude of the flow in the simulations agree
perfectly with the PIV results.\\
In experiments it is found that a secondary air effect, the
``surface seal", has a significant influence on the cavity shape
at high Froude number. Since the surface seal phenomenon (and its
influence) is more pronounced at higher impact velocities, it
limits our experimental Froude number range. In the boundary
integral simulations the air was intentionally excluded,
thus avoiding this limitation.\\

Because the velocity of the impacting disk is a constant control
parameter in our experiments, a simple theoretical argument based
on the collapse of an infinite, hollow cylinder describes the key
aspects of the transient cavity shape.\\

This model accurately reproduces the dynamics of the cavity
including its maximal expansion and total collapse time. It also
captures the scaling for the depth of closure and the total depth
of the cavity at pinch--off, and predicts their ratio to be close
to 2, where 2.1 is found in experiments and simulation. \\
There is a close similarity of this description to the cavity
dynamics proposed by \cite{Duc06}. However, by introducing the
asymmetry between the radial expansion and collapse, we find a
better agreement between the theory and the radial dynamics of the
cavity. The fact that the flow is qualitatively different during
expansion of the cavity on the one hand and its contraction on the
other is found to be responsible for the asymmetry. Our approach
is also conceptually different, as \cite{Duc06} take
$\alpha_{expa}$ to be independent of the Froude number, while we
allow it to be weakly dependent on Froude and, more importantly,
our description includes the last stage of the collapse, which
is solely driven by inertia.\\

We find the volume of air entrained by the impact of the disk to
behave as $V_{bubble}/h_0^3 \propto (1 +
0.26\text{Fr}^{1/2})\text{Fr}^{1/2}$. This dependence is set by
the Froude dependence of two length scales, namely  the axial
length scale, distance between the pinch-off point and the disk,
and the radial expansion of the cavity. Here we have excellent
agreement between the experimental and numerical findings and the
prediction of the model.\\

Finally, the appendices deal with the time evolution of the cavity
radius we discussed previously in \cite{ber06}. In this paper, and
subsequent papers dealing with the universality of the last stages
of the pinch-off from a theoretical point of view (\cite{gor06} and
\cite{egg07}), the pinch-off is assumed to be symmetric around the
closure depth, whereas in experiment and numerics we observe that
the minimal radius of the void actually moves downward in time. As
this (small) axial translation could be relevant for this
universality issue, it is studied in detail in
Appendix~\ref{app:minimal radius}, where we find that it can be
included within the model presented in this paper, as a secondary
effect.


\begin{acknowledgments}
{\it Acknowledgements} The work is part of the research  program of
FOM, which is financially supported by NWO. R.B. and S.G.
acknowledge financial support.
\end{acknowledgments}\

\begin{appendix}

\section{Revisiting \cite{ber06}} \label{appendixA}

In this appendix we will review the results which were presented in
\cite{ber06} as far as they are necessary for the understanding of
the material in Appendix~\ref{app:minimal radius}, together with
some additional results.\\

\subsection{Neck radius at closure depth}\label{sec:rad_closure}

The most prominent length scale describing the cavity dynamics
close to pinch-off is the neck radius. It can be taken either at
its minimum (see also Appendix~\ref{app:minimal radius}) or at the
constant depth of closure at each instant of time. In this
Appendix we will deal with the latter, closely following the discussion
in our earlier letter (\cite{ber06}).\\

\begin{figure}
\includegraphics[width=1\textwidth]{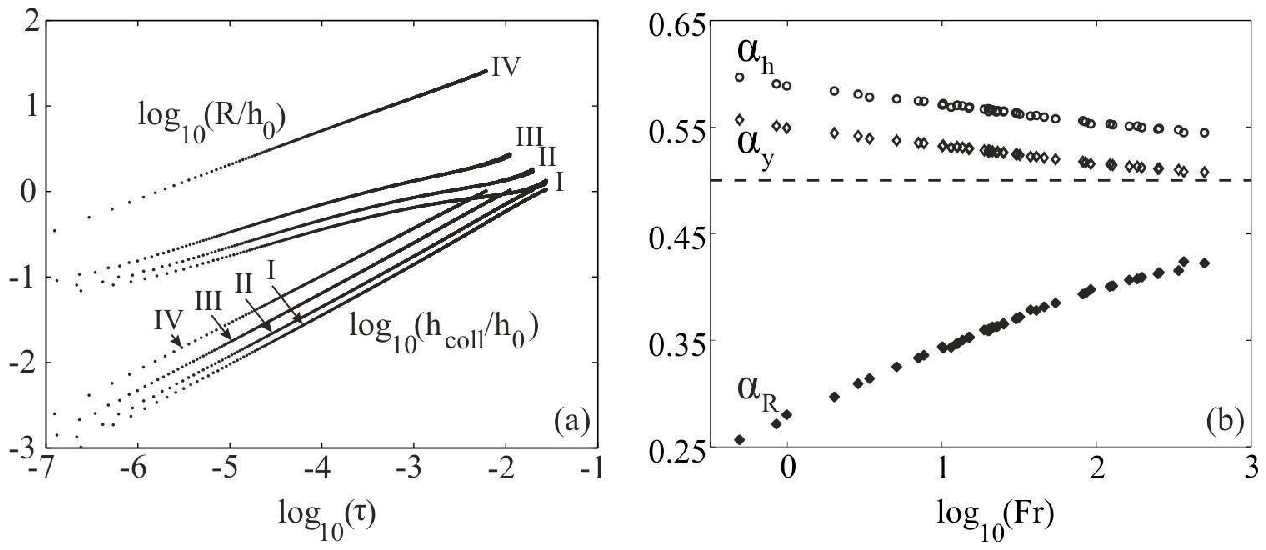}
\caption{(a) Time-evolution of the neck radius at closure depth
$h_{coll}(tau)$ and the radius of curvature $R(\tau)$ in the axial
direction for $\text{Fr}=3.4$, $5.1$, $10.2$, and $200$. Both are
found to follow a power law with respective exponents $\alpha_h$ and
$\alpha_R$. (b) Power-law exponents $\alpha_h$ and $\alpha_R$ of the
neck radius and radius of curvature as a function of the Froude
number. Here we have added the exponents $\alpha_{y}$ which are
corrected for the logarithmic factor in Eq.~(\ref{eq:neckradius}).
In the large $\text{Fr}$ limit, all exponents approach the
Rayleigh-value of {1/2}
[cf.~Eq.~(\ref{eq:traj3})].}\label{fig:app_radius}
\end{figure}

The neck radius $h_{coll}$ over the time $\tau$ is found to obey a
power law scaling (see Fig.~\ref{fig:app_radius}a), where the
exponent is observed to vary between 0.55-0.62, depending on the
Froude number (Fig.~\ref{fig:app_radius}b). So, for all Froude
numbers the scaling exponents are above the value of 1/2 that is
expected from Eq.~(\ref{eq:continuity}). This result is consistent
with the recent careful experiments of \cite{tho07}, for the
somewhat different experiment in which an air bubble pinches off
from an underwater nozzle.

The deviation can be partly understood by considering the full
two-dimensional Rayleigh-type equation Eq.~(\ref{eq_ray1a}) instead
of only the first term as was done in the derivation of
Eq.~(\ref{eq:traj3}). The procedure is described in \cite{ber06} and
\cite{gor05} and introduces a logarithmic correction to the neck
radius,
\begin{equation}\label{eq:neckradius}
h_{coll}(t)\cdot\left(-\log(h_{coll})\right)^{1/4}\sim\tau^{1/2} \,
.
\end{equation}
However, even though this result improves the description of the
experiments and numerics, small deviations are still seen in the
dynamics of $h_{coll}(t)$ for small Froude numbers. These
deviations, which depend on the Froude number, show the influence of
the initial conditions on the early stage of the pinch-off.\\

As was described in \cite{ber06}, these deviations suggest that the
neck radius is not the only relevant length scale for the cavity
shape around the pinch-off point. As the cavity shape in axial
direction is approximately parabolic, the second characteristic
length scale can conveniently be chosen as the radius of curvature
$R$ in the $z$ direction, which is defined as \beq
\label{eq:parabolicb} 1/R(t)=d^2r/dz^2|_{z=z_{min}}. \eeq We found
the time dependence of the radius of curvature to also follow a
power-law with a Froude-dependent exponent $\alpha_R$. The scaling
exponents of the neck radius and the radius of curvature differ
significantly from one another at small Froude number, but tend to
converge to $1/2$ for higher
$\text{Fr}$ (see Fig.~\ref{fig:app_radius}a and b).\\

\subsection{Cavity shape at pinch-off} \label{sec:neck radius}

If the pinch off would be self-similar, the free-surface profiles
near the closure point at different times (Fig.~\ref{fig:collapse
profiles}a) would superpose when rescaled by any one
characteristic length, e.g., the neck radius $h(\tau)$. Due to the
different time-dependence of the neck radius and the radius of
curvature, however, such an operation fails to collapse the void
profiles onto a single shape. To characterize the free-surface
shapes, one thus needs to consider the neck radius $h_{coll}(t)$
and the radius of
curvature $R(t)$.\\

\begin{figure}
\begin{center}
\includegraphics[width=0.6\textwidth]{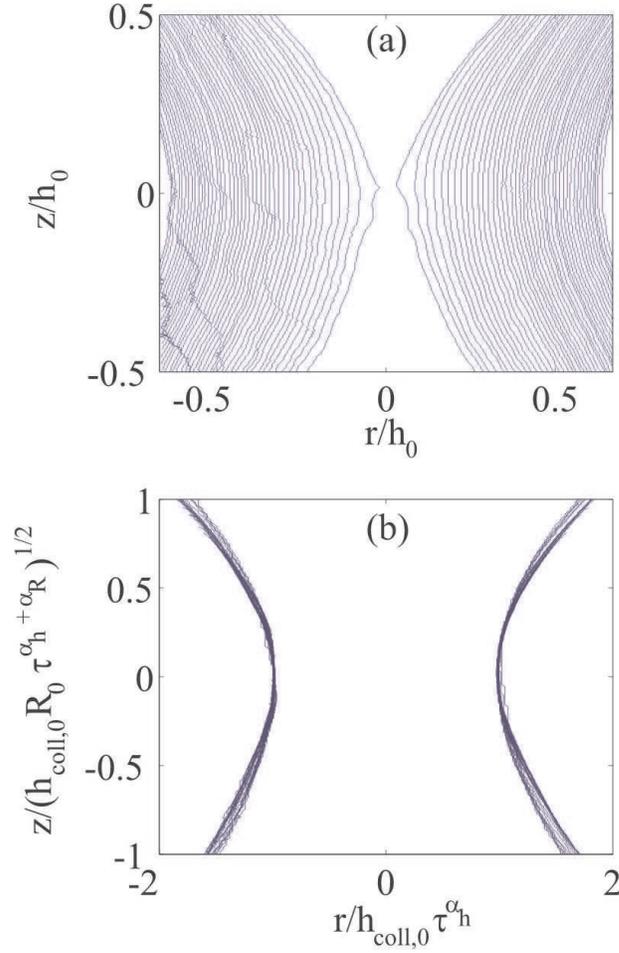}
\caption{The profiles of the void obtained by experiments for $h_0
=$ 30 mm and $V = 1.0$ m/s (Fr $= 3.4$). (a) Void profile obtained
at different instances in time. (b) Void profile in which the radial
and axial coordinates were rescaled with the powerlaws of
$h_{coll}(t)$ and $\sqrt{h_{coll}(t)R(t)}$ respectively. Here, the
numerically determined power--laws were used for the neck radius
$h_{coll}(\tau)$ and the radius of curvature $R(\tau)$ (see main
text). } \label{fig:collapse profiles}
\end{center}
\end{figure}

As was argued in \cite{ber06}, if the radial dimension $r$ is scaled
by $h_{coll}$, it follows from the locally parabolic shape and
Eq.~(\ref{eq:parabolicb}) that the axial dimension $z$ should be
scaled by $\sqrt{h_{coll}R}$. Instead of using the actual length
scales to rescale the profiles in Fig.~\ref{fig:collapse profiles}b,
we use the power laws that were obtained from the numerical
simulations to collapse all experimental profiles onto a single
curve. This signals once more the excellent agreement between the
simulations and the experiment, as the scaling exponents are
obtained from the simulations and the profiles from experiment.

As will be discussed in Appendix \ref{app:minimal radius}, the
axial position of the minimum radius $z_{min}$ moves down somewhat
as the collapse progresses and it is therefore necessary to
translate the profiles in the vertical direction to match the
position of the minimum radius. For the inertially driven pinch
off of a bubble from an underwater nozzle \cite{tho07} also find
that two length scales are necessary to collapse the free surface
profiles.

Recent theoretical calculations by \cite{gor06} and \cite{egg07} for
the symmetric inertial pinch off of a single bubble (either
initiated by surface tension or a straining flow) indicate that the
scaling law for $h_{coll}(\tau)$ has an exponent which slowly varies
with time, i.e., strictly speaking it does not scale. Our findings
so far cannot confirm or disprove this theory, since our experiments
and boundary integral simulations do not have sufficient temporal
range to find the small deviations in the exponent. To escape the
limitation of the experimental range set by the viscosity, surface
tension, and air, the experiment should be scaled up to an
unrealistic size (with a container size of over $10^3$
meters\footnote{The current experiments, in a $0.5 \times 0.5 \times
1$ m$^3$ container, cover 2 orders of magnitude in $h(\tau)$ whereas
$10/2 = 5$ orders of magnitude are needed.}) to match the needed
precision of at least 10 decades in time
presented in the numerical calculations of \cite{egg07}.\\

\section{Minimal neck radius} \label{app:minimal radius}

\begin{figure}
\begin{center}
\includegraphics[width=1\textwidth]{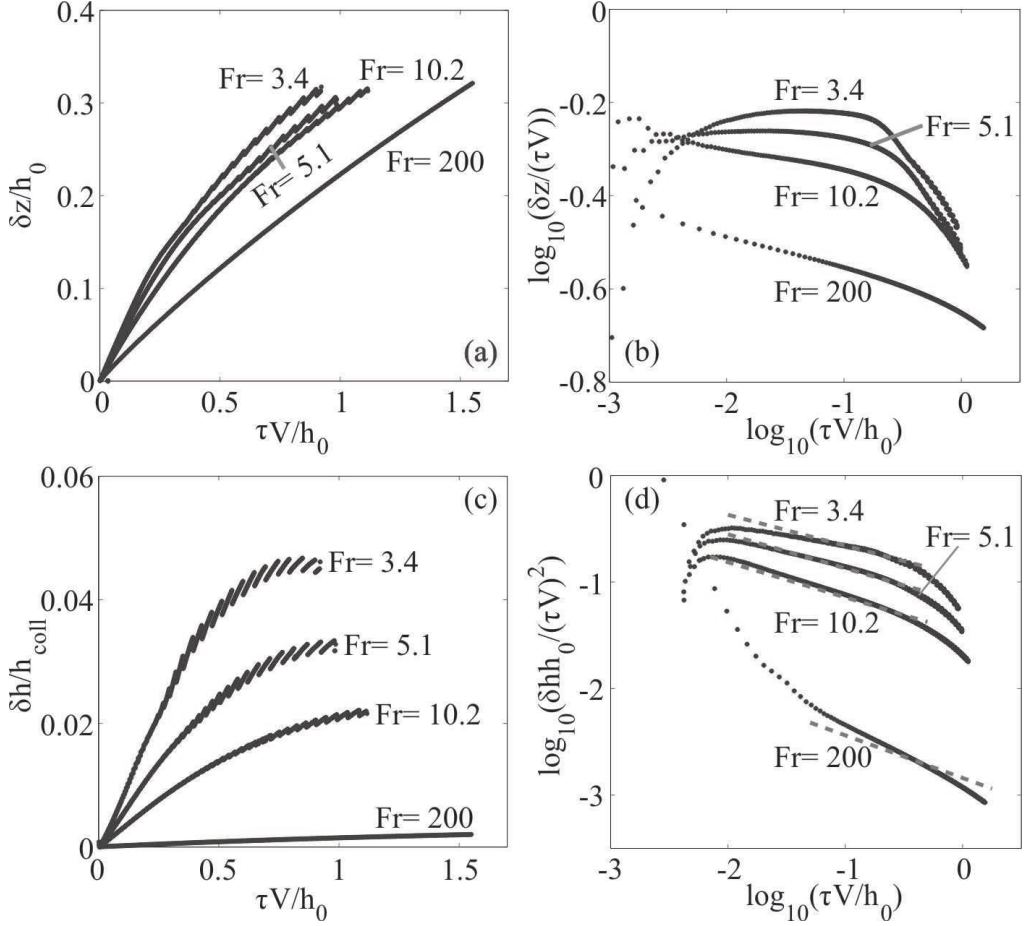}
\caption{(a) The difference $\delta z(\tau)$ between the depth of
the minimal radius $z_{min}(\tau)$ and the depth of closure
$z_{coll}$ (both normalized by the disk radius $h_0$) as a function
of the time interval $\tau$ (normalized by $h_0/V$) remaining until
pinch--off for different Froude numbers. Each data series is
obtained from the boundary integral simulations and starts when the
minimal radius equaled the disk radius $h_0$. (b) Doubly logarithmic
plot of the same data as (a), only now $\delta z$ is compensated by
$\tau$ revealing an interval in which $\delta z$ is proportional to
$\tau$ (showing as a horizontal line in the plot). (c) The
difference $\delta h(\tau) = h_{coll}(\tau) - h_{min}(\tau)$ of the
radius at closure depth $h_{coll}(\tau)$ and the minimal radius
$h_{min}(\tau)$ of the cavity, normalized by $h_{coll}(\tau)$, as a
function of $\tau$ (normalized by $h_0/V$) for different Froude
numbers. Clearly, $\delta h/h_{coll}$ is smaller for increasing
Froude number which results from the fact that the cavity shape
becomes more cylindrical for high $\text{Fr}$. (d) Doubly
logarithmic plot of the same data as in (c), only now $h_{coll} -
h_{min}$ is compensated by $\tau^2$ to reveal the resultant scaling
$\tau^{-\alpha _R}$ (see text). The dashed red lines are the scaling
exponents $\alpha_R(\text{Fr})$ found by determining the time
evolution of the radius of curvature in the axial direction through
the procedure outlined in \cite{ber06} ($\alpha_{R}(\text{Fr=3.4}) =
0.29; \alpha_{R}(\text{Fr=5.1}) = 0.32; \alpha_{R}(\text{Fr=10.2}) =
0.33$; and $\alpha_{R}(\text{Fr=200}) = 0.40)$). The jumps in the
data of Figures (a) and (c) are a result of the regridding routine
in our boundary integral simulation and have no physical meaning.}
\label{fig:z_min}
\end{center}
\end{figure}

Recently, a lot of attention has been given to the time evolution of
the minimal cavity radius $h_{min}(\tau)$ at the closure depth and
the (non-)universality of its scaling behavior approaching the
pinch-off, see \cite{gor05}, \cite{ber06}, \cite{gor06}, and
\cite{egg07} (cf. also Appendix~\ref{appendixA}). Especially in the
theoretical analysis of \cite{gor06} and \cite{egg07} it is a key
assumption that the dynamics is symmetric around the pinch-off
point. In this Appendix we will show that in our setup this
assumption is \emph{not} satisfied, as the position of
the minimum neck radius actually moves downwards in time.\\

In Figure~\ref{fig:z_min}a the difference $\delta z(\tau)$ between
the depth at which the cavity radius is minimal $z_{min}(\tau)$
and the depth of closure $z_{coll}$ is plotted as a function of
the time interval $\tau$ remaining until pinch--off for four
different Froude numbers. Each data series is obtained from the
boundary integral simulations and, as we focus on the behavior
close to pinch--off, starts when the minimal radius is equal to
the disk radius $h_0$.

We observe that the dynamics near pinch--off can be reasonably
well described by a simple proportionality $\delta z \propto
\tau$. This becomes even more clear if we compensate $\delta z$ by
$\tau$ in a double logarithmic plot of the same data
(Fig.~\ref{fig:z_min}b), which reveals a plateau in which $\delta
z/\tau$ becomes independent of time, especially for the lower
three Froude numbers.\\

As the pinch-off is approached, not only the depth of the minimal
radius $z_{min}(\tau)$ converges to the depth of closure
$z_{coll}$, but naturally also the minimal radius itself
$h_{min}(\tau)$ approaches the radius at the depth of closure
$h_{coll}(\tau)$ as can be seen in Fig.~\ref{fig:z_min}c. The
relative difference $\delta h(\tau)$ between $h_{min}$ and
$h_{coll}$ is seen to be smaller for increasing Froude number due
to the cavity taking a more cylindrical shape at higher Froude
number.

Since the cavity profile is locally parabolic close to the
pinch--off point the approach of $h_{min}$ to $h_{coll}$ is
described as: \beq \delta h(\tau)=\frac{[\delta
z(\tau)]^2}{2R(\tau)}\,, \label{eq:geometry}\eeq where $R(\tau)$
is the radius of curvature in the axial direction at the minimum
neck radius.

This radius of curvature was found to exhibit power-law scaling in
time with a Froude-dependent exponent $R(\tau) \propto
\tau^{\alpha_R}$ (cf. \cite{ber06} and Appendix~\ref{appendixA},
Fig.~\ref{fig:app_radius})
which we can combine with the linear time dynamics found for
$\delta z$ in Fig.~\ref{fig:z_min}. 
This leads to $h_{min}$ approaching $h_{coll}$ as \beq \delta h
\propto \tau^{2-\alpha_R}. \eeq Indeed, in Fig.~\ref{fig:z_min}d
this scaling is confirmed for the lower Froude numbers. For the
highest Froude number ($\text{Fr}=200$) the data seem to deviate
from this scaling due to small deviations from $\delta z \propto
\tau$ which are observed for this Froude number in
Fig.~\ref{fig:z_min}b.\\

The final question we want to address in this section is whether
it is possible to understand the relation $\delta z \propto \tau$
from the model. We start from Eq.~(\ref{eq:traj3}) with
$t_{coll}=\Delta t$ given by Eq.~(\ref{eq:closure time}), i.e.,
for $t_{cross}<t<t_{coll}$
\begin{equation}
h(z,t) = \sqrt{2h_0\alpha_{ctra}V}\sqrt{\Delta t
\small{(z)}\,\,-\,\,t\,\,} \,.
\end{equation}
To find the depth of minimal radius we now have to compute the
derivative of $h(z,t)$ to $z$ and equate to zero, or, equivalently
\begin{eqnarray}
\frac{\partial }{\partial z}[h(z,t)^2] = 2h_0\alpha_{ctra}V
\left[\frac{\partial \Delta t}{\partial z}\right] &=& 0
\qquad\Rightarrow \\ \frac{1}{V} -
(\alpha_{expa}\beta_{expa}+\alpha_{ctra}\beta_{ctra})\frac{h_0V}{gz^2}
&=& 0\,,
\end{eqnarray}
which leads to the conclusion that the depth of the minimal radius
is always equal to the closure depth, i.e., independent of time
(cf. Eq.~(\ref{eq:closure depth})). Clearly this is in
disagreement with the observations. This was to be expected as the
translation of the depth of the minimal radius is quite small,
typically an order of magnitude smaller than other length scales,
e.g., the closure depth). We will therefore have to look for a
second order effect.\\

To take the discussion one step further, we return to
Fig.~\ref{fig:sqrt}a, where we find that $\alpha_{ctra}$ slightly
decreases with increasing Froude number. From this we can infer
(at least qualitatively) that $\alpha_{ctra}$ also very slightly
decreases with depth. This stands to reason, as $\alpha_{ctra}$
shows the same trend (in $\text{Fr}$) as $\alpha_{expa}$, which is
the ratio of the initial expansion velocity of the cavity and the
disk velocity. As at greater depth the hydrostatic pressure is
larger it is expected that $\alpha_{expa}$ should decrease with
depth.

Now letting $\alpha_{ctra}$ (and the other parameters
$\alpha_{expa}$, $\beta_{expa}$, and $\beta_{ctra}$) depend on $z$
we have
\begin{equation}
\frac{\partial }{\partial z}[h(z,t)^2] =
2h_0V\left(\frac{\partial\alpha_{ctra}}{\partial z}\right)\,\tau +
2h_0V\alpha_{ctra}\small{(z)}\left(\frac{\partial \Delta
t\small{(z)}}{\partial z}\right) = 0 \,.
\end{equation}
For any fixed Froude number $\text{Fr}_0$ we can now Taylor expand
$\partial \Delta t/\partial z$ around the closure depth $z_{coll}$
which leads to
\begin{equation}
\frac{\partial\Delta t}{\partial z} = \left(\frac{\partial\Delta
t}{\partial z}\right)_{z_{coll}} +\,\,\, (\delta
z)\left(\frac{\partial^2\Delta t}{\partial z^2}\right)_{z_{coll}}
= \,\,\,(\delta z)\left(\frac{\partial^2\Delta t}{\partial
z^2}\right)_{z_{coll}} \,,
\end{equation}
as within the model the closure depth is defined by the condition
$(\partial \Delta t/\partial z)_{z_{coll}} = 0$ (see
subsection~\ref{sec:closure depth}). Recall that $\delta z =
z-z_{coll}$. Combining the above two equations then leads to the
following relation between $\delta z$ and $\tau$
\begin{equation}\label{eq:delzvstau}
\delta
z=-\left[\frac{1}{\alpha_{ctra}}\left(\frac{\partial\alpha_{ctra}}{\partial
z}\right)\left(\frac{\partial^2\Delta t}{\partial
z^2}\right)^{-1}\right]_{z_{coll}}\tau \,,
\end{equation}
where it is good to note that $\Delta t$ does not only directly
depend on $z$, but also indirectly, through the parameters
$\alpha_{ctra}$, $\alpha_{expa}$, $\beta_{expa}$, and
$\beta_{ctra}$.

From the data presented in Fig.~\ref{fig:sqrt}a we can perform a
local second order fit to $\alpha_{ctra}$ as a function of
$\sqrt{\text{Fr}}$ around $\sqrt{\text{Fr}_0}$. Assuming that the
relation between closure depth and Froude number also holds near
the closure depth, i.e., $z/h_0 = C_1 \sqrt{\text{Fr}}$ with $C_1
\approx 1.10$ (see again subsection~\ref{sec:closure depth}) we
can translate this into a quadratic expression in $\delta z$
\begin{equation}
\alpha_{ctra} \approx \alpha_{ctra,\,0} + k_1 \frac{\delta z}{h_0}
+ k_2 \frac{(\delta z)^2}{h_0^2} \,,
\end{equation}
and similarly for the other parameters $\alpha_{expa}$,
$\beta_{expa}$, and $\beta_{ctra}$.

As an example we take $\text{Fr}_0 = 5.1$ for which we find
$\alpha_{ctra,\,0} = 0.56$, $k_1 = -0.087$, and $k_2 = 0.038$.
After a straightforward but quite lengthy calculation we evaluated
the prefactor in Eq.~(\ref{eq:delzvstau}) to give $\delta z/h_0
\approx 0.20 \tau V/h_0$. This is in the same direction and of the
same order of magnitude as the result from our boundary integral
simulation where we find $\delta z/h_0 \approx 0.51 \tau V/h_0$
(see Fig.~\ref{fig:z_min}a and b). Repeating this procedure for
$\text{Fr} = 10.2$ $\text{Fr} = 200$ yields proportionality
constants of $0.16$ and $0.074$ respectively, which correctly
predict the downward trend with increasing Froude number that is
also observed in the numerical simulation, where we find $0.42$
and $0.25$ respectively. Considering the large number of
approximations made in this calculation and the subtlety of the
effect the agreement is remarkable.





\end{appendix}

\end{document}